\documentclass[aps,prb,twocolumn,showpacs,floatfix]{revtex4}
\usepackage{graphicx}

\newcommand{\gen}[1]{
#1}
\newcommand{\blue}[1]{
#1}
\newcommand{\green}[1]{
#1}
\newcommand{\magenta}[1]{
#1}

\newcommand{\last}[1]{\emph{#1}}

\newcommand{\etc}{{\it etc.}\ }
\newcommand{\etal}{{\it et al.}\ }
\newcommand{\REBCO}[1]{#1Ba$_2$Cu$_3$O$_{7-\delta}\ $}
\newcommand{\YBCO}{\REBCO{Y}}

\newcommand{\SC}{su\-per\-con\-ducting\ }
\newcommand{\ZPB}{Z.\ Phys.\ B\ }
\newcommand{\ML}{Mater.\ Lett.\ }
\newcommand{\PRB}{Phys.\ Rev.\ B\ }
\newcommand{\PRL}{Phys.\ Rev.\ Lett.\ }

\newcommand{\APL}{Appl.\ Phys.\ Lett.\ }
\newcommand{\JAP}{J.\ Appl.\ Phys.\ }
\newcommand{\JMR}{J. Mater. Res.\ }
\newcommand{\PC}{Physica\ C\ }
\newcommand{\SUST}{Super\-cond.\ Sci.\ Tech\-nol.\ }
\newcommand{\EPL}{Euro\-phys.\ Lett.\ }
\newcommand{\MSE}{Mater.\ Sci.\ Eng.\ }
\newcommand{\Nat}{Nature\ }
\newcommand{\EQ}[1]{Eq.\,(\ref{#1})}

\newcommand{\FIG}[1]{Fig.\,\ref{#1}}

\newcommand{\REF}[1]{Ref.\,\onlinecite{#1}}
\newcommand{\TBL}[1]{Table~\ref{#1}}
\newcommand{\BE}{\begin{equation}}
\newcommand{\BEA}{\begin{eqnarray}}
\newcommand{\EE}{\end{equation}}
\newcommand{\EEA}{\end{eqnarray}}
\newcommand{\Xmm}{-5mm}  

\begin{document}

\title[Inter- and intra\-grain currents]%
{Inter- and intra\-grain currents in bulk melt-grown YBaCuO rings}

\author{A.\,B.\,Surzhenko}
\thanks{The author to whom a correspondence
should be addressed} \altaffiliation[On leave from ]{Institute for
Magnetism, Kiev, Ukraine} \email{surzhenko@ipht-jena.de}
\author{M.\,Zeisberger}
\author{T.\,Habisreuther}
\author{W.\,Gawalek}
\affiliation{Institut f\"ur Physikalische Hochtechnologie,
Winzerlaer Str.\,10, Jena D-07745, Germany}
\author{L.\,S.\,Uspenskaya}
\affiliation{Institute of Solid State Physics, Russian Academy of
Science, Institutskii pr.\,16, Chernogolovka, Moscow district,
142432, Russia}

\begin{abstract}
A simple contactless method suitable to discern between the
inter\-grain (circular) current, which flows in the thin
super\-conducting ring, and the intra\-grain current, which does
not cross the weakest link, has been proposed. At first, we show
that the inter\-grain current may directly be estimated from the
magnetic flux density $B(\pm z_0)$ measured by the Hall sensor
positioned in the special points $\pm z_0$ above/below the ring
center. The experimental and the numerical techniques to determine
the value $z_0$ are discussed. Being very promising for
characterization of a current flowing across the joints in welded
YBaCuO rings (its dependencies on the temperature and the external
magnetic field as well as the time dissipation), the approach has
been applied to study corresponding properties of the intra- and
inter\-grain currents flowing across the $a$-twisted grain
boundaries which are frequent in bulk melt-textured YBaCuO
samples. We present experimental data related to the flux
penetration inside a bore of MT YBaCuO rings both in the
non-magnetized, virgin state and during the field reversal. The
shielding properties and their dependence on external magnetic
fields are also studied. Besides, we consider the flux creep
effects and their influence on the current re-distribution during
a dwell.
\end{abstract}


\date{\today}
\maketitle

\section{\label{sec:Intro} Introduction}

Unique feature of thin \SC rings has to be well-known to everybody
who ever tried to investigate the inter\-grain
currents flowing across weak links in poly\-crystalline \YBCO
(YBCO) ceramics and/or thin
films~\cite{Leiderer88,Mohamed91,Jung93,Darhmaoui96}. By changing
the outer $R_o$ and the inner $R_i$ radii of the ring, one may
essentially enlarge a difference between the length scales,
$R_o+R_i$ and $R_o-R_i$, over which, respectively, the inter- and
the intra\-grain currents flow. Provided that the ring is thin
enough $(R_o-R_i)\ll(R_o+R_i)$, one may reliably register the
magnetic flux density $B$ induced by the inter\-grain, shielding
current even if the intra\-grain currents much exceed it.

The melt-textured (MT) growing process~\cite{Jin88} has generally
allowed to escape an appearance of the large-angle grain
boundaries (GBs) and, thereby, the current localization inside the
grains. For this reason, a worldwide interest to the ring-like
geometry was fading away until an idea to weld~\cite{Salama92} MT
blocks gained a
respectable reputation~\cite{Shi95,%
Philip98,Zheng99,Delamare00,Zheng01,Prikhna01,Noudem01,Harnois01,Puig01,%
Walter01,Yoshioka02,Kord01,Claus01,Kambara02}. Briefly, the
joining procedures consist in welding of single MT domains during
a liquid-assisted process which either releases residual
BaCuO-oxide trapped in the MT material~\cite{Philip98} or uses
corresponding REBCO (RE=Tm~\cite{Zheng99,Zheng01,Prikhna01},
Yb~\cite{Delamare00,Noudem01}, Er~\cite{Walter01,Yoshioka02},
Y~\cite{Shi95,Kambara02} or Y+Ag~\cite{Harnois01,Puig01,Claus01})
compounds melting at lower temperatures. Since these techniques
were shown to produce the joints capable to carry high currents,
such artificial joining opens new perspectives for fabrication of
large-scale \SC devices (e.g., magnetic bearings, electro\-motors
and generators, energy storage systems, \etc\cite{Murakami00}).
Their performance, though, crucially depends on the density $j_w$
of the critical current which the weld may transmit.

Despite of an obvious necessity to compare the joints obtained
under different welding conditions, surprisingly few authors
reported numerical values $j_w$. Besides, most of these estimates
was obtained by the resistive measurements which, because of the
ohmic losses in the current pads, are severely limited by the
maximum admissible current $I_w=j_wS\alt10^3\,A$. Thus, these
measurements are restricted both by the surface $S$ of the weld
and by its quality $j_w$ which, in turn, imposes the lowest
temperature margin. On the other hand, prevalent contact\-less
methods (e.g., levitation force technique~\cite{Zheng99,Kord01},
magneto-optical image
analysis~\cite{Zheng99,Walter01,Pannetier01}, the scanning
Hall-sensor magneto\-metry~\cite{Philip98,Zheng99,Delamare00,Zheng01,%
Prikhna01,Harnois01,Puig01,Yoshioka02} and magne\-ti\-za\-tion
loop studies~\cite{Philip98,Zheng99,Claus01}) are not free of
ambiguities in processing of experimental data unless, as
mentioned above, the welded sample has a shape of thin ring. In
this case, the circular current throughout a ring is limited by
the weakest link which is, in turn, usually associated with a
joint.

Among the mentioned above experimental techniques, the scanning
Hall sensor magneto\-metry looks preferable. The main reason is
that, having no restrictions for a size of joints, this method is
quite suitable for characterization of large-scale welds required
for practical applications. The same feature seems also useful for
other tasks to be eventually resolved. In particular, one has to
study a degradation of the critical current densities from a center of large
MT domains to their rims~\cite{Dewhurst98} and its relation to a
content of the precursor mixture, the growth temperature, \etc
Aside from a knowledge of the best growth conditions, such studies
will allow to determine the optimum size of welded parts. The
last, but not the least argument is a widespread use of the
scanning Hall sensor systems%
~\cite{Philip98,Zheng99,Delamare00,Zheng01,Prikhna01,Harnois01,%
Puig01,Yoshioka02}.

For this technique to give reliable results, the experimental data
have to be properly processed. Meanwhile, most of the authors
prefer to simplify the genuine distribution of currents and to
consider their rings as \emph{homogeneous} (one-turn) solenoids
carrying the transport current $I_w$. In other words, they neglect
the intra\-grain currents, i.e. those which do not cross the
joint. We shall hereafter demonstrate that such simplification is
applicable only for rough estimates. More accurate method, i.e. a
scan of the flux trapped by a ring with its subsequent computer
fit~\cite{Zheng01}, is also far from perfect. The problem is that
a necessity to move the Hall sensor makes this technique hard to
exploit at different temperatures and/or external magnetic fields.
Moreover, the scanning duration $t_{sc}$ of, typically, a few
minutes during which the currents may noticeably dissipate,
assents to start the measurements only after a long-continued
dwell $t_{dw}\gg t_{sc}$. So, an important information about the
current losses appears unavailable.

Section~\ref{sec:Model} of this manuscript represents simple
analytical equations constituting a backbone of the novel
experimental approach which combines the advantages of both
methods and is free of their faults. We show that to find out the
densities of the inter- and the intra\-grain currents one does not
need to waste a time on the magnetic flux scan. Actually, these
densities may, with a proper accuracy, be estimated from two
experimental points $B(r,z)$ measured at certain distances $z$
above/below the ring center. In Section~\ref{sec:Results} we apply
this method to study the current distribution in the large-scale
MT YBCO rings. The results are summarized in
Section~\ref{sec:Summary}.

\section{\label{sec:Model} Model and its solution}

\begin{figure}[!b] \begin{center}
\includegraphics[angle=-90,width=0.56\columnwidth]{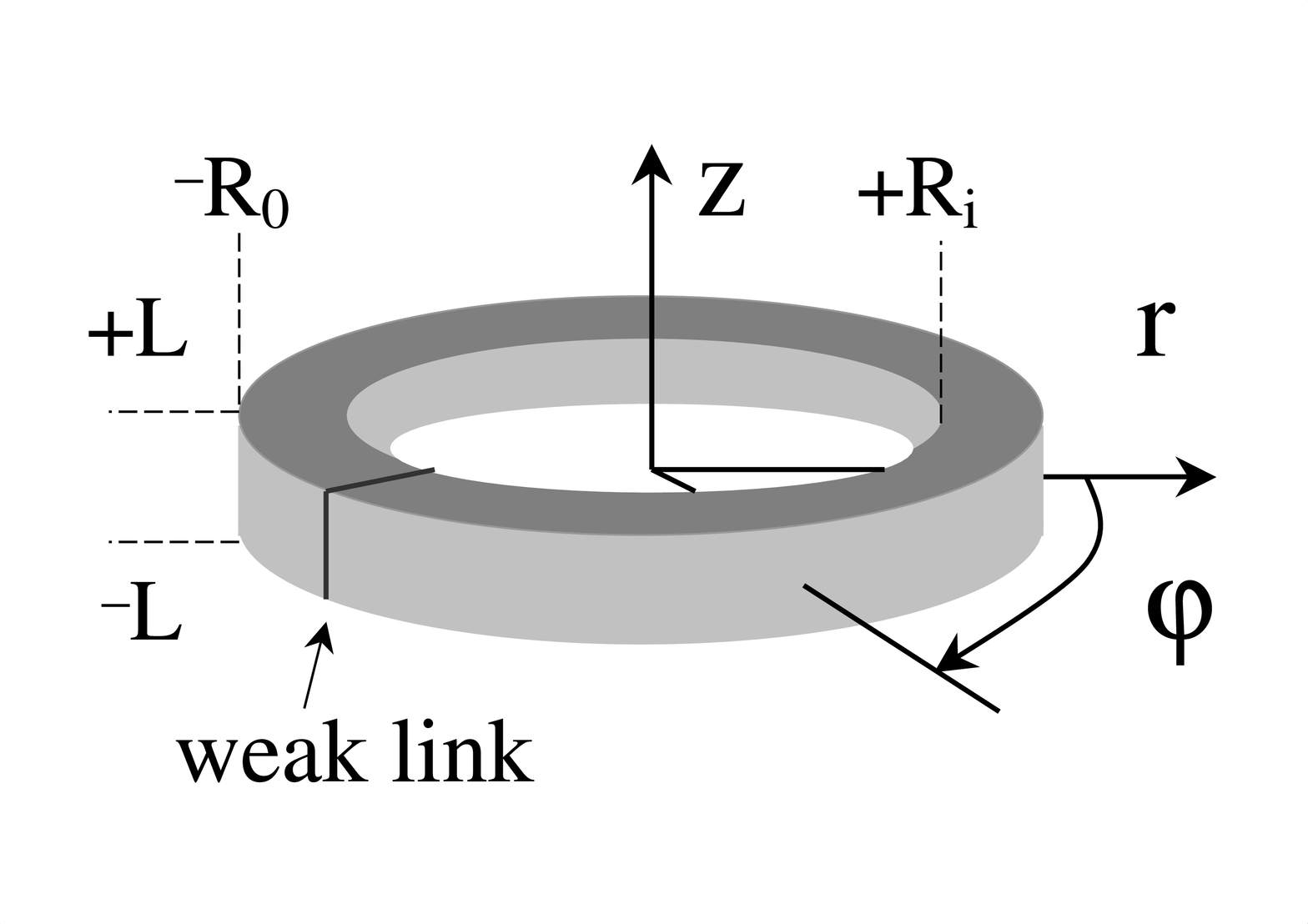}
\end{center}\vspace{\Xmm}
\caption{Super\-conducting ring which contains weak link.}
\label{ring}
\end{figure}

Let us consider a homogeneous ring (see \FIG{ring}) made of \SC
material with the critical current density $j_m$ and introduce
therein a weak link transmitting a current up to
\BE I_w=j_w(R_o-R_i)2L=j_m f (R_o-R_i)2L \label{Iw} \EE
where $f=j_w/j_m<1$. It is worth of mentioning that the weak link
makes this ring \emph{in\-homo\-geneous} with respect to a
rotation around its axis, $z$. Let the ring was completely
magnetized in strong magnetic fields applied along the
$z$-direction. The subject of our interest will be the currents
flowing in the ring as well as the radial profiles $B(z=Const,r)$
of \blue{the vertical component} of the magnetic flux density
which these currents produce.

\begin{figure}[!tb] \begin{center}
\includegraphics[width=0.58\columnwidth]{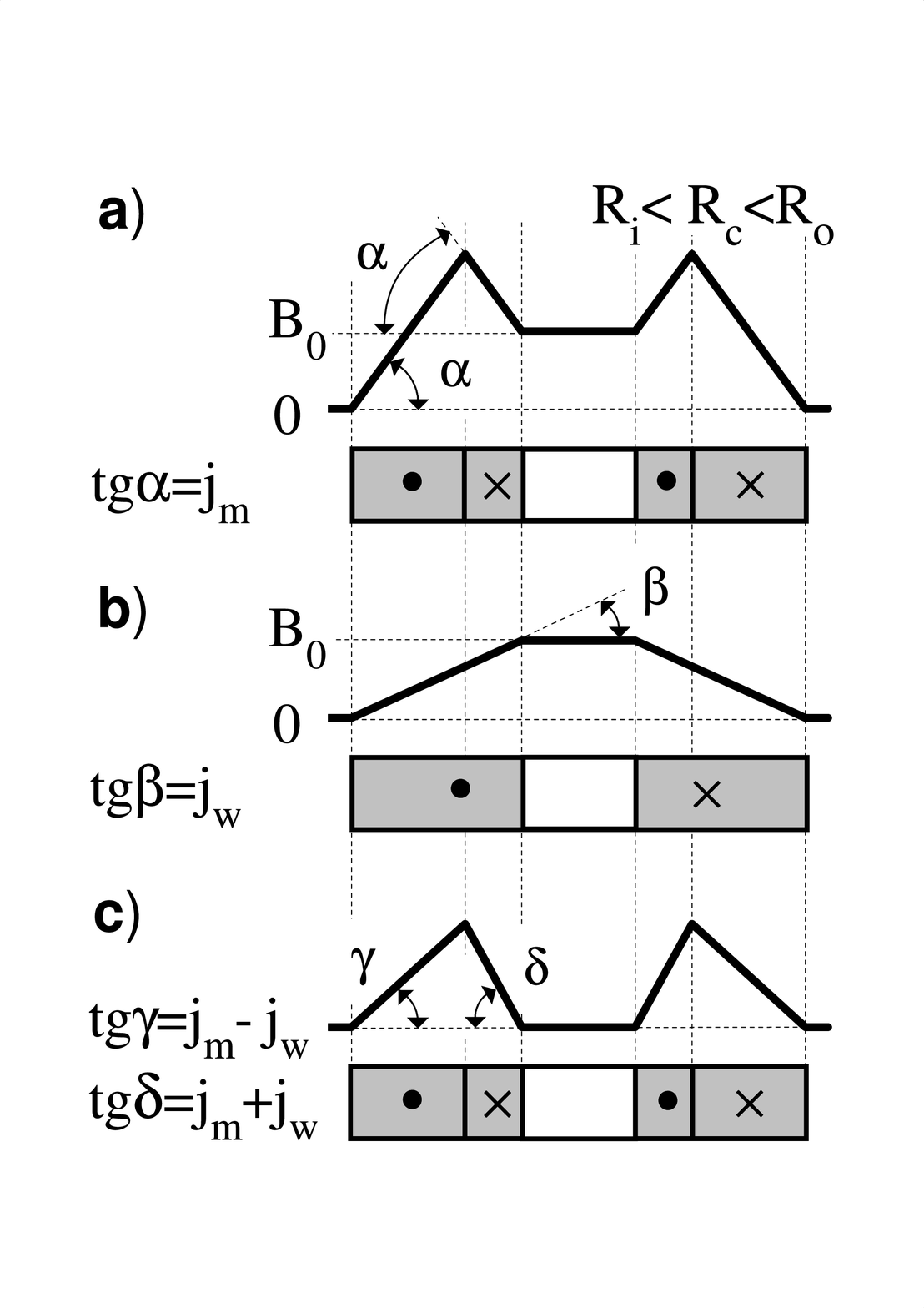}
\end{center}\vspace{\Xmm}
\caption{Profiles of the density $B$ of the remanent magnetic flux
trapped by a ring containing weak links a) and its components,
$B_w$ and $B_m$, induced by b) the inter- and c) the intra\-grain
currents, respectively.} \label{profiles}
\end{figure}

Assuming the Bean's critical state model \blue{in the simplest
form of infinitely high ($L\rightarrow \infty$) sample} to be
valid, one can readily outline the profile passing through the
point $r=0$ and the ring area which is far enough from the weak
link. \FIG{profiles}a shows this profile in the remanent state
($H=0$). Of necessity, this situation can readily be expanded to
the case of arbitrary fields by adding a constant background
$\mu_0H$. For this reason, we shall thenceforward take into
account only two components of the magnetic flux, $B_w$ and $B_m$.
These are induced by the inter- ($I_w$) and the intra\-grain
($I_m$) currents which, respectively, crosses and does not cross
the weak link. \blue{When \blue{$L\rightarrow \infty$}}, the
former results to the triangle profile truncated in the ring bore
by a plateau $B_0=j_w(R_o-R_i)$ (\FIG{profiles}b), whereas the
latter responds for the difference $B_m=B-B_w$ (see
\FIG{profiles}c). The radius $R_c$ wherein the intra\-grain
current alters its direction from the clockwise (CW) to the
counter\-clockwise (CCW) is known
\BE R_c(f)=\left[R_o+R_i-f(R_o-R_i)\right]/2\label{rc} \EE
to change between $(R_i+R_o)/2$ at $f=0$ and $R_i$ when
$f=1$~\cite{Zheng01}. Using \EQ{rc}, one can calculate the
amplitude
\BE I_m=j_m(R_o-R_i)L(1-f^2) \label{Im} \EE
and demonstrate that the CW and CCW intra\-grain currents
compensate each another, i.e. the current loop $I_m$ does appear
closed inside the \SC grain.

\begin{figure}[!tb] \begin{center}
\includegraphics[width=0.75\columnwidth]{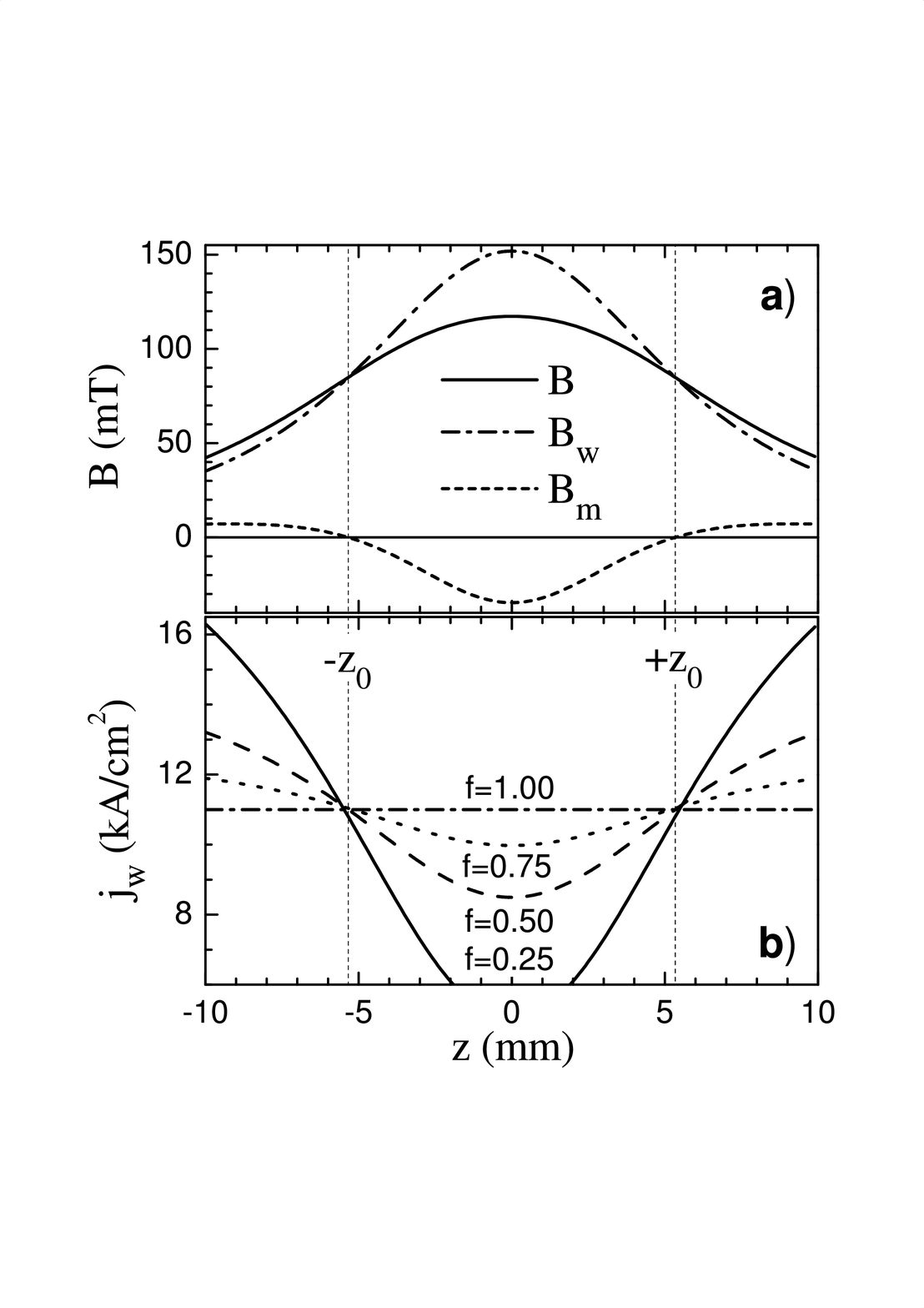}
\end{center}\vspace{\Xmm}
\caption{Figure a) shows how the densities of the remanent
magnetic flux $B$ and its components, $B_w$ and $B_m$, along the
axis ($r=0$) of the ring depend on the distance $z$ from its
center, $z=0$. The ring sizes, $R_o=10\,mm$, $R_i=5.5\,mm$,
$2L=3.8\,mm$, as well as $j_w=11\,kA/cm^2$ are borrowed from the
work of Zheng \etal~\cite{Zheng01}. The missing parameter
$f=j_w/j_m$ is taken to be $f=0.5$ ($j_m=22\,kA/cm^2$). The $j_w$
vs $z$ curves in figure b) are calculated from the total magnetic
flux $B$ within the one-turn approximation (\ref{BT}) and, thus,
represent an error appearing when the term $B_m$ is neglected. }
\label{example}
\end{figure}

In the case \blue{$L\rightarrow \infty$ illustrated in
\FIG{profiles}}, $B_m(r<R_i)\equiv 0$, i.e. any point inside the
ring bore is suitable to estimate the density $j_w=B_0/(R_o-R_i)$.
This comfortable situation is, though, too far from reality. Since
the finite-\blue{size} rings are not free of de\-magnetizing
effects, the term $B_m(r<R_i)$ is actually non-zero. Thus, in
order to estimate the current $I_w$ and/or its density $j_w$, one
has to calculate and to remove the component $B_m$ from the
experimentally accessible value $B(z,r)$. In this paper we
\blue{solve this problem} in a bit another way. We show and gladly
exploit that $B_m(z,r)$ is an alternating function: it is negative
inside a certain region surrounding the ring center ($z=0,r=0$)
and positive outside it (see, for example, \FIG{example}a). Thus,
positioning the Hall sensor somewhere at the border
$B_m(z_0,r_0)=0$, one can directly measure the \blue{inter\-grain}
flux component $B_w(z_0,r_0)$ and use this value to restore a
``pure'' circular current $I_w$ flowing throughout a ring.

One could, certainly, search for a whole surface which satisfy the
condition $B_m(z_0,r_0)=0$. But let, for simplicity, restrict our
search by the points along the ring axis, $r_0=0$. The magnetic
flux density of thin ($r_i=r_o=R$), one-turn solenoid at $r=0$ is
well-known
\BEA
B=\frac{I}{4L}\left[\frac{L-z}{\sqrt{R^2+(L-z)^2}}+\frac{L+z}%
{\sqrt{R^2+(L+z)^2}}\right]. \label{thin} \EEA
\blue{For finite-size rings, one requires} to average the function
(\ref{thin}) over the range between their outer $r_o$ and the
inner $r_i$ radii. Using a formal procedure \blue{$B=\int
B(r)\,dr/\int dr$}, one has \blue{a general solution}
\BEA B&=&j\cdot\Phi(L,z,r_o,r_i).\label{thick} \EEA
Here $j$ denotes the current density $j=I/[2L(r_o-r_i)]$ and the
function $\Phi$ gives the effective size depending on four
distances
\BEA \Phi(L,z,r_o,r_i)&=&\Gamma(L-z,r_o,r_i)+\Gamma(L+z,r_o,r_i),
\nonumber\\ \Gamma(x,r_o,r_i)&=&\frac{x}{2}
\ln{\left(\frac{r_o+\sqrt{r_o^2+x^2}}{r_i+\sqrt{r_i^2+x^2}}\right)}.
\label{F} \EEA
Using these equations, one can obtain the magnetic flux densities,
\blue{$B_w$, $B_m$ and $B=B_w+B_m$}, produced by each of three
magnetic systems shown in \blue{\FIG{profiles} for the case of
finite sizes. For example, the coil (\FIG{profiles}b), which
carries a current $I_w$ of a density $j_w$, yields
\BEA B_w=j_w\Phi(L,z,R_o,R_i). \label{BT} \EEA
The intra\-grain flux component $B_m$ is created by by a pair of
coils (see \FIG{profiles}c) having the same height $2L$ and the
common radius $R_c$ given by \EQ{rc}. Since these coils carry a
current $I_m$ of densities $j_m(1-f)$ and $-j_m(1+f)$, one
has}\magenta{
\BEA B_m&=&K\cdot j_m[(1-f)\Phi(L,z,R_o,R_c(f))-\nonumber\\
&-&(1+f)\Phi(L,z,R_c(f),R_i)]\label{BL} \EEA
where $K$ is a geometrical factor which origin is illustrated in
\FIG{diamond}. One can see that a weak link is surrounded by the
diamond-shaped area (marked by bright color) where the
intra\-grain current $I_m$ flows no more in the circular direction
(as \EQ{thick} assumes), but either toward or outward the ring
center. Owing to such an arrangement, $I_m$ within this
``diamond'' gives a negligible contribution to $B(z,r=0)$. Then,
the flux density $B_m$ of a full intra\-grain loop has to be
corrected onto a factor $K\simeq 1-\delta/2\pi R_c$ (see
\FIG{diamond}) where the half-width $\delta$ may be estimated as
the length necessary to transmit a whole current $I_m$ given by
\EQ{Im}. Since the radial current density $j_r$ can not exceed
$\sqrt{j_m^{\,2}-j_w^{\,2}}$ (see vector diagrams in
\FIG{diamond}), one can obtain
\begin{figure}[!tb] \begin{center}
\includegraphics[angle=-90,width=0.65\columnwidth]{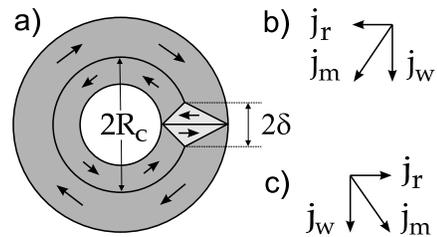}
\end{center}\vspace{\Xmm}
\caption{a) Currents which flow in a fully-magnetized
super\-con\-ducting ring with a weak link and vector diagrams
explaining their distribution inside the diamond-shaped area b)
above and c) below a weak link.} \label{diamond}
\end{figure}
    \BEA \delta=\frac{1}{2}(R_o-R_i)\sqrt{1-f^2}. \EEA
and, then, calculate the parameter
    \BEA K=1-N\frac{R_o-R_i}{4\pi R_c(f)}\sqrt{1-f^2} \label{K} \EEA
for the case of several weak links, $N$. This approximation is no
longer valid at $N\delta> \pi R_c$, i.e. when the diamonds start
to overlap each other.}

In order to imagine how the flux densities vary along the ring
axis, we inserted parameters of the real ring reported in
\REF{Zheng01} to Eqs.(\ref{BT}) and (\ref{BL}) and presented these
curves in \FIG{example}a. These data once more convince us that
the one-turn approximation (\ref{BT}) may properly be applied to
experimental values $B(z)$ only when $B_m=0$, i.e. at the points
$z_0\approx \pm 5.4\,mm$ on the given axis ($r=0$). Otherwise,
such method may considerably underestimate ($|z|<z_0$) or
overestimate ($|z|>z_0$) the current density $j_w$. For the
parameter $f=0.5$ in \FIG{example}a, the $j_w$ value, respectively
of the Hall sensor position $0\leq |z|\leq 10\,mm$, changes in the
wide range ($77\ldots 120\%$) around the genuine density
$j_w=11\,kA/cm^2$. This error essentially increases when
$f\rightarrow 0$ (see \FIG{example}b).

\begin{figure}[!tb] \begin{center}
\includegraphics[width=0.85\columnwidth]{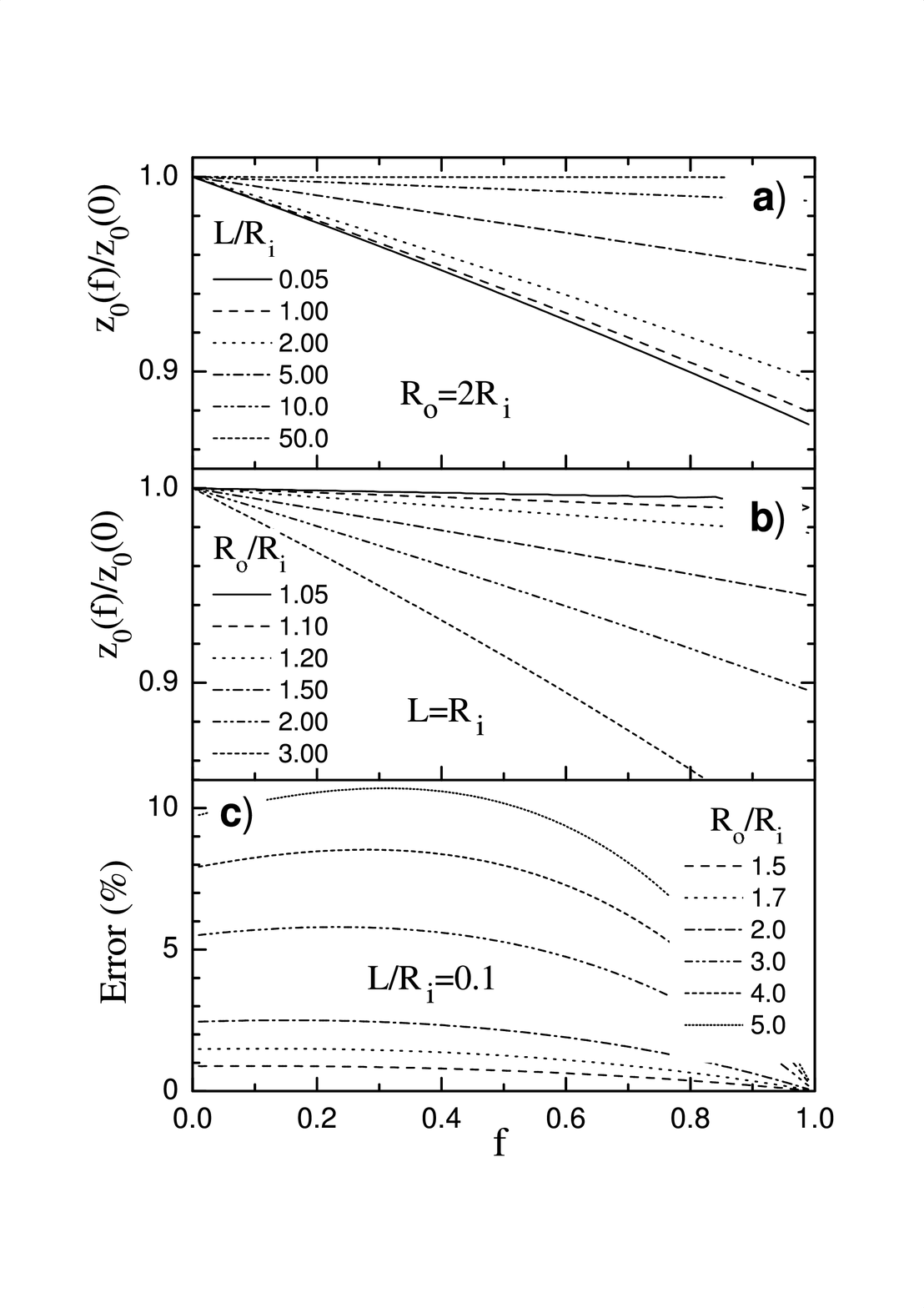}
\end{center}\vspace{\Xmm}
\caption{a), b) The $z_0$ vs $f$ dependencies calculated from
\EQ{BL=0} at various parameters $R_o$, $R_i$ and $L$. Figure c)
presents the error $B_m/B_w(z_0)$ which appear in the worse case
of thick, flat rings ($R_o/R_i>1.5$, $L/R_i=0.1$) provided that
the $z_0$ vs $f$ dependence is neglected, namely,
$z_0(f)=z_0(0)$.} \label{z_vs_f}
\end{figure}

In order to escape such situation, one has to \blue{know} $z_0$.
The main problem, which will encounter everybody trying to do it,
is that the expression inside the brackets in \EQ{BL} depends on
the unknown factor $f=j_w/j_m$  and so, certainly, does $z_0$
which satisfies
\BEA
\frac{\Phi(L,z_0,R_o,R_c(f))}{\Phi(L,z_0,R_c(f),R_i)}%
=\frac{1+f}{1-f}.\label{BL=0} \EEA
If one could ignore this dependence, $z_0$ would be defined only
by the ring geometry and, hence, readily calculated.

Upon a closer view at \FIG{example}b, one can really note that
$z_0$ tends to the ring center as $f$ increases. However, the
effect seems so weak that we did not resist a temptation to know
in which rings its disregard still gives an acceptable error. We
studied the $z_0$ vs $f$ curves at various parameters $R_o$, $R_i$
and $2L$ and revealed almost no deviation from the initial value
$z_0(f=0)$ in thin ($R_i\rightarrow R_o$) and/or high
($L\rightarrow \infty$) rings (see \FIG{z_vs_f}a,b), i.e. in those
which are close to a form of ideal sole\-noid. As the largest
deviation is, therefore, expected in thick, flat rings ($R_o\gg
R_i$, $L\ll R_i$), it is the case to be inspected. So, we
calculated for such rings the amendment $B_m/B_w|_{z_0}$ which
has, by the definition, to be a zero unless the $z_0$ vs $f$
dependence is ignored. The points $z_0(R_i,R_o,L)=Const$, wherein
this ratio was estimated, were taken from \EQ{BL=0} reduced by the
substitution $f=0$ to rather comfortable form~\footnote{Numerical
methods (e.g., the bi\-sectional search) determine the roots $\pm
z_0$ of \EQ{BL(f=0)=0} to any desired accuracy. For this reason,
we desisted from attempts to reduce \EQ{BL(f=0)=0} into an
absolute form $z_0=f(R_i,R_o,L)$ which can be obtained by
introducing further restrictions for the ring geometry.}
\BEA \Phi(L,z_0,R_o,R_c(0))=\Phi(L,z_0,R_c(0),R_i),
\label{BL(f=0)=0} \EEA
where $R_c(0)=(R_o+R_i)/2$. Accepting the admissible error to be
2.5\% and restricting, thereby (see \FIG{z_vs_f}c), the studied
rings by not too stringent condition~\footnote{Owing to a finite
height of real rings, a requirement for the ratio $R_o/R_i$ is
actually yet less stringent.} $R_o/R_i\leq 2$, we are qualified to
use the above simplification, $z_0(f)=z_0(0)$, in practice.
\begin{figure}[!b] \begin{center}
\includegraphics[angle=-90,width=0.86\columnwidth]{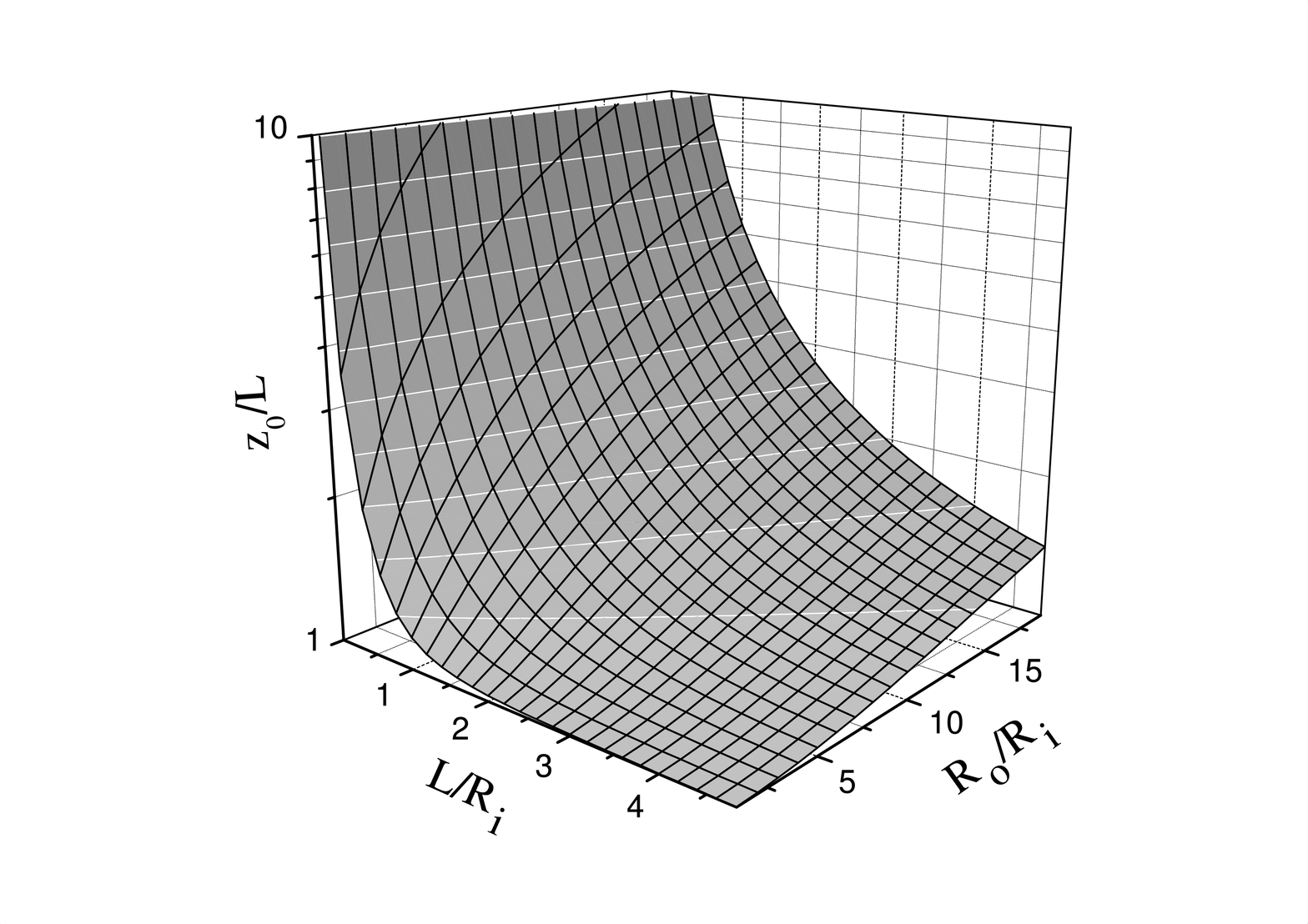}
\end{center}\vspace{\Xmm}
\caption{The solutions of Eq.\,(\ref{BL(f=0)=0}) normalized to the
ring half-height. Note that everywhere, except for the case of an
ideal solenoid ($L\rightarrow\infty$, $R_o/R_i\rightarrow 1$),
$z_0$ appears outside of the ring, $|z_0|\geq L$. Thus, the method
is not restricted by the bore diameter $2R_i$ which may be smaller
than the size of the Hall sensor holder.} \label{z0}
\end{figure}

Thus, knowing the ring sizes ($R_o$, $R_i$ and $2L$) only, one
may, on the one hand, pre-determine $z_0$ from \EQ{BL(f=0)=0} by
numerical methods. Such solutions (normalized, for the sake of
convenience, to the ring half-height $L$) are shown in \FIG{z0}.
On the other hand, this numerical procedure (at least, in the case
of welded rings) may \emph{optionally} be avoided. Before joining,
these rings inevitably contain one or more slits which prevent a
circular current to flow ($f=0$). So, the points $\pm z_0$,
wherein the rema\-nent flux density of the \emph{cut} ring changes
its sign (see \FIG{example}a), are available for direct
measurements. Then, registering $B(\pm z_0)$ above or below the
center of the \emph{welded} sample, one may readily calculate the
joint quality
\BEA j_w=B(z_0)/\Phi(L,z_0,R_o,R_i).\label{jw} \EEA
Since $z_0$ depends only on the ring geometry, there is no matter
whether the welded ring has the same $j_m$ as that in the original
material or not. This feature does look important because heating
of the MT samples up to temperatures close to their melting
points, which actually is the joining procedure, often results to
the oxygen losses and requires to re-oxygenate the samples.

\section{\label{sec:Samples} Experimental details}

The initial MT YBCO samples, from which we cut the rings, were
grown by the top-seeding method which is described in details
elsewhere~\cite{Doris}. Briefly, commercially purchased powders of
YBa$_2$Cu$_3$O$_{7-\delta}$ and Y$_2$O$_3$ were thoroughly mixed
in proportions providing the final phase composition
Y:Ba:Cu=1.5:2:3 and diluted by 1\,wt.\% CeO$_2$. This mixture was
uni\-axially pressed into cylindrical pellets which sizes were
selected to exceed the sizes of future rings. Then, the self-made
SmBa$_2$Cu$_3$O$_X$ seed was installed in the center of each
pellet and oriented so that its crystallo\-graphic axis $c$,
[001], was parallel to the axis of cylinder. On the morrow of
iso\-thermal melt-growth ($T=988^\circ$C) the samples were cooled
down to the room temperature and then post-annealed in flowing
oxygen.

The rings were carved (by using the aluminium oxide grind\-stones)
so that the $z$-axis of each ring was parallel to the $c$-axis of
the initial MT crystal.

The larger half of experiments was in scanning of the
\emph{remanent} flux distribution. These scans were performed with
the 3D-positioning system~\cite{Klupsch} allowing to move the Hall
sensor step\-wise with a pitch of 100\,$\mu m$ in the horizontal
plane and of 75\,$\mu m$ along the vertical direction. The
standard procedure was the following. The rings were mounted
inside a small vessel and field-cooled (FC) in magnetic fields
$\mu_0H\approx 1\,T$ to the liquid nitrogen temperature
$T=77.3\,K$. Then, the field was switched off (with a sweep rate
$0.5\,T/min$) and a ring (still inside a vessel with a liquid
nitrogen) was moved to an area available for a scan. To eliminate
the undesirable scatter because of a dissipation of the remanent
flux during the scanning time (typically, of a few minutes), a
scan was always run after a long-continued dwell
($t_{dw}=30\,min$). Both the axial $B(r=0,z)$ and the radial
$B(r,z=Const)$ profiles were studied.

To measure the hysteresis loops $B(H)$ as well as the flux
dissipation $B(t)$, we mounted the Hall sensor directly on the
studied ring. The Hall sensor positioning error did not exceed
$50\,\mu m$.

The magnetic moment, $m$, was registered by the commercial VSM
(Model No.\,3001, Oxford Instriments Ltd). Owing to a relatively
large distance between the vibrating ring and the pick-up coils,
this value may roughly be associated with $B(z=\infty)$.

The radial flux distribution at the ring edges $z=\pm L$ in small
magnetic fields ($H<3\,kOe$) was also visualized by using a
high-resolution magneto-optical imaging
technique~\cite{Uspenskaya97,Uspenskaya03}.

\section{\label{sec:Results} Results and discussion}

In this section we report extensive experimental studies of the
inter- and the intra\-grain currents for the ring-like geometry.
At first, we investigate the current distribution in the remanent
state and compare our results with those obtained by
con\-ventional methods. Then, we study a behavior of rings
containing natural weak links (i.e. the boundaries between the
$a$-twisted grains) in external magnetic fields. Finally, we
explore the flux creep effects.

\subsection{\label{subsec:cut_rings} Cut rings in the remanent state}

It is obvious that the fewer fitting parameters are included in an
equation, the more trustworthy one can test whether it fits an
experiment or not. For this reason, we start our studies from the
cut rings wherein relevant is the only parameter, $j_m$.

When a ring contains \magenta{$N$ slits} and transmits, hence, no
\gen{inter\-granular current $I_w$}\,, the magnetic flux density
along the ring axis ($r=0$) is given by \EQ{BL} at $f=0$\magenta{
\BEA B=K j_m[\Phi(L,z,R_o,R_c(0))-\Phi(L,z,R_c(0),R_i)],\
\label{B-Cut} \EEA
where $R_c(0)$ is the average radius, $(R_o+R_i)/2$, and
\BEA K=1-\frac{N}{2\pi}\cdot\frac{R_o-R_i}{R_o+R_i}\ \label{K0}
\EEA}
One has to note that the term in brackets totally defines the
shape of the $B$ vs $z$ curve, i.e. $j_m$ may change only its
amplitude. Thus, estimating a disagreement between theoretical and
experimental profiles $B(r=0,z)$, one can readily check to what
extent may our approach be relied on. In particular, one can put
the solutions $z_0$ of \EQ{BL(f=0)=0} to the experimental proof.
By changing the ring sizes, we can also test how $j_m$ varies with
a distance from the seeding point both in the axial and in the
radial directions.

\begin{table}[!b]\vspace{\Xmm}
\caption{Theoretical and experimental values $z_0$ in the cut
rings (CR) \magenta{consisting of two 180$^\circ$-arcs ($N=2$).
The} estimates for the intra\-grain current density $j_m$ are also
included.} \label{tbl:z0}
\begin{ruledtabular}
\begin{tabular*}
{\hsize}{@{\extracolsep{0ptplus1fil}}ccccccc@{\extracolsep{0pt}}}

Ring\ & \multicolumn{3}{c}{Ring sizes\footnote{\ The ring sizes
are given with the accuracy $\pm 0.03\,mm$.},\,$mm$} &
\multicolumn{2}{c}{$z_0,\,mm$} & $j_m$\footnote{\ The $j_m$ values
are obtained in the self-field (remanent state) after the
long-continued dwell, $t_{dw}=30\,min$.},\\

No. & $R_o$ & $R_i$ & $2L$ & theory & exp. & $kA/cm^2$\\
\colrule
   CR1 & 6.05 & 2.95 & 5.35 & 3.92 & 3.70 &  \magenta{17.9$_8$}
\\ CR2 & ''   & ''   & 4.90 & 3.83 & 3.67 &  \magenta{18.9$_8$}
\\ CR3 & ''   & ''   & 3.98 & 3.60 & 3.52 &  \magenta{20.2$_3$}
\\ CR4 & ''   & ''   & 3.30 & 3.46 & 3.40 &  \magenta{21.3$_4$}
\\ CR5 & ''   & ''   & 2.81 & 3.37 & 3.34 &  \magenta{22.6$_2$}
\\ CR6 & ''   & ''   & 2.22 & 3.27 & 3.23 &  \magenta{22.8$_3$}
\\ CR7 & ''   & ''   & 1.55 & 3.20 & 3.21 &  \magenta{22.9$_7$}
\\ CR8 & 4.99 & ''   & ''   & 2.89 & 2.90 &  \magenta{22.2$_3$}

\end{tabular*}
\end{ruledtabular}
\end{table}

\begin{figure}[!t] \begin{center}
\includegraphics[angle=-90,width=0.98\columnwidth]{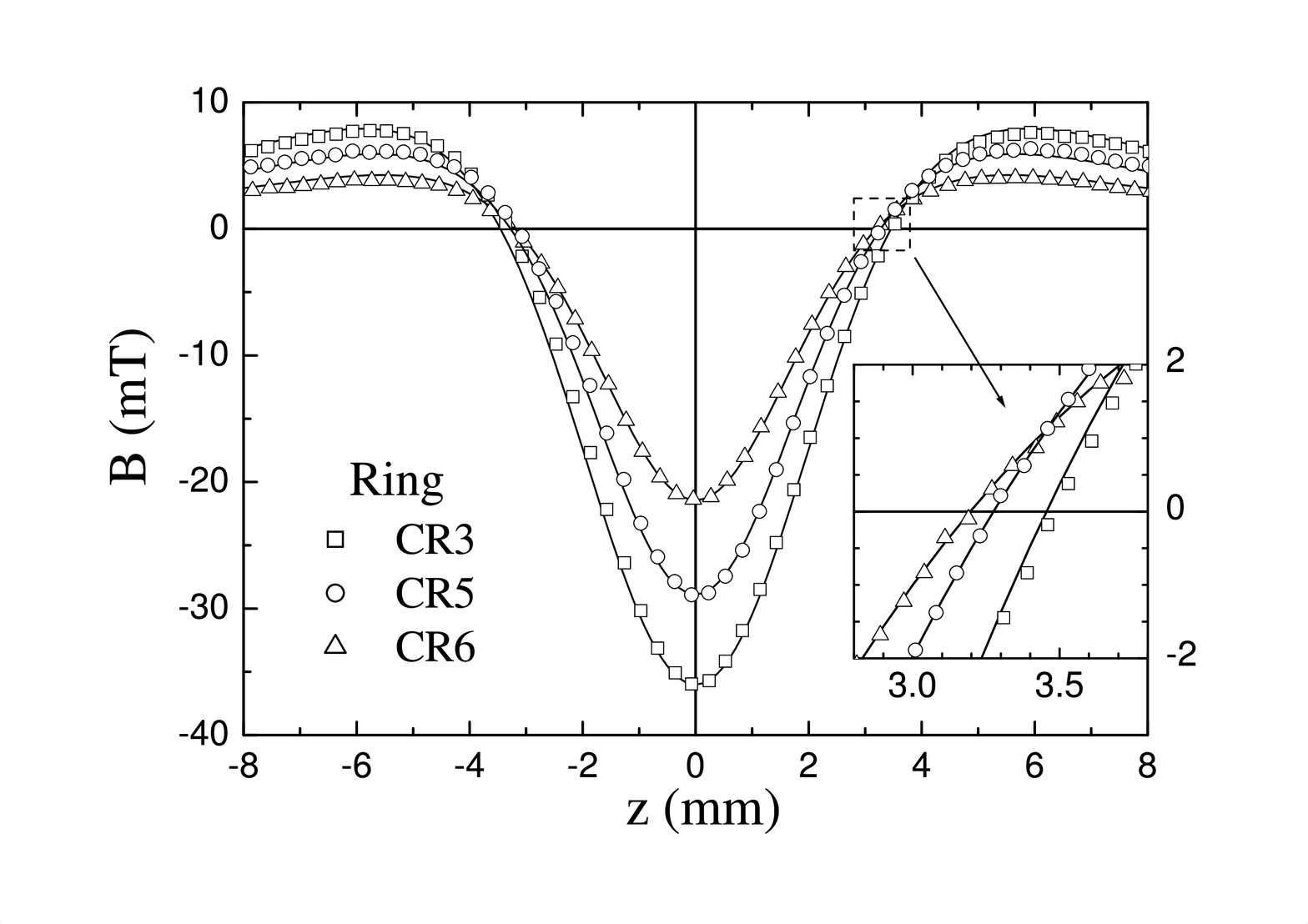}
\end{center}\vspace{\Xmm}
\caption{The axial profiles $B(r=0,z)$ of the remanent flux
density (open points) measured at $T=77.3\,K$ along the axis of
the cut rings (see details in \TBL{tbl:z0}) and their
approximations (solid curves) given by \EQ{B-Cut}. Inset magnifies
the area around the points $B(z_0)=0$.}
\label{cutrings}\vspace{\Xmm}
\end{figure}

For these studies we used the ring (see \TBL{tbl:z0}) which was
cut from the central part of the MT crystal, i.e. the ring center
($r=0$) coincided with the seeding point. The inner diameter,
$2R_i=5.90\,mm$, was selected to exceed the size ($\approx 5\,mm$)
of the Hall sensor holder and to make, therefore, the ring
suitable for the $z$-scan along the whole axis $r=0$. At first, we
cut this ring onto two 180$^\circ$-arcs and glued them together by
using an epoxy resin. A tough\-ness of such mechanical contact was
good enough for a succeeding treatment of the ring by
grind\-stones. Then, grinding, step-by-step, the ring bottom (i.e.
reducing the ring height), we measured at each stage the axial
profile $B(r=0,z)$ of the remanent flux density. During the
measurements we exactly followed the procedure described in
Section\,\ref{sec:Samples}. Some of these data are presented in
\FIG{cutrings} (open symbols). To calculate $j_m$ and to recover
from \EQ{B-Cut} theoretical approximations (solid curves), we used
the $B$-points near the ring center, $z=0$, wherein the flux
density has the maximum amplitude. \FIG{cutrings} shows that
\EQ{B-Cut} well fits the experimental data. However, more careful
look (see \TBL{tbl:z0}) reveals a certain difference. One can
notice that the genuine value $z_0$ (estimated as the
half-distance between the points wherein the experimental curve
intersects the $X$-axis) in high rings is usually a bit smaller
than $z_0$ predicted by \EQ{BL(f=0)=0}. This feature is
accompanied by the obvious tendency for $j_m$ to decrease (from
the top surface to the bottom one) and may readily be explained by
the growth-related in\-homogeneities of the MT
material~\cite{Diko00,Surzhenko02b}.

\begin{figure}[!b] \begin{center}\vspace{\Xmm}
\includegraphics[angle=-90,width=0.4\columnwidth]{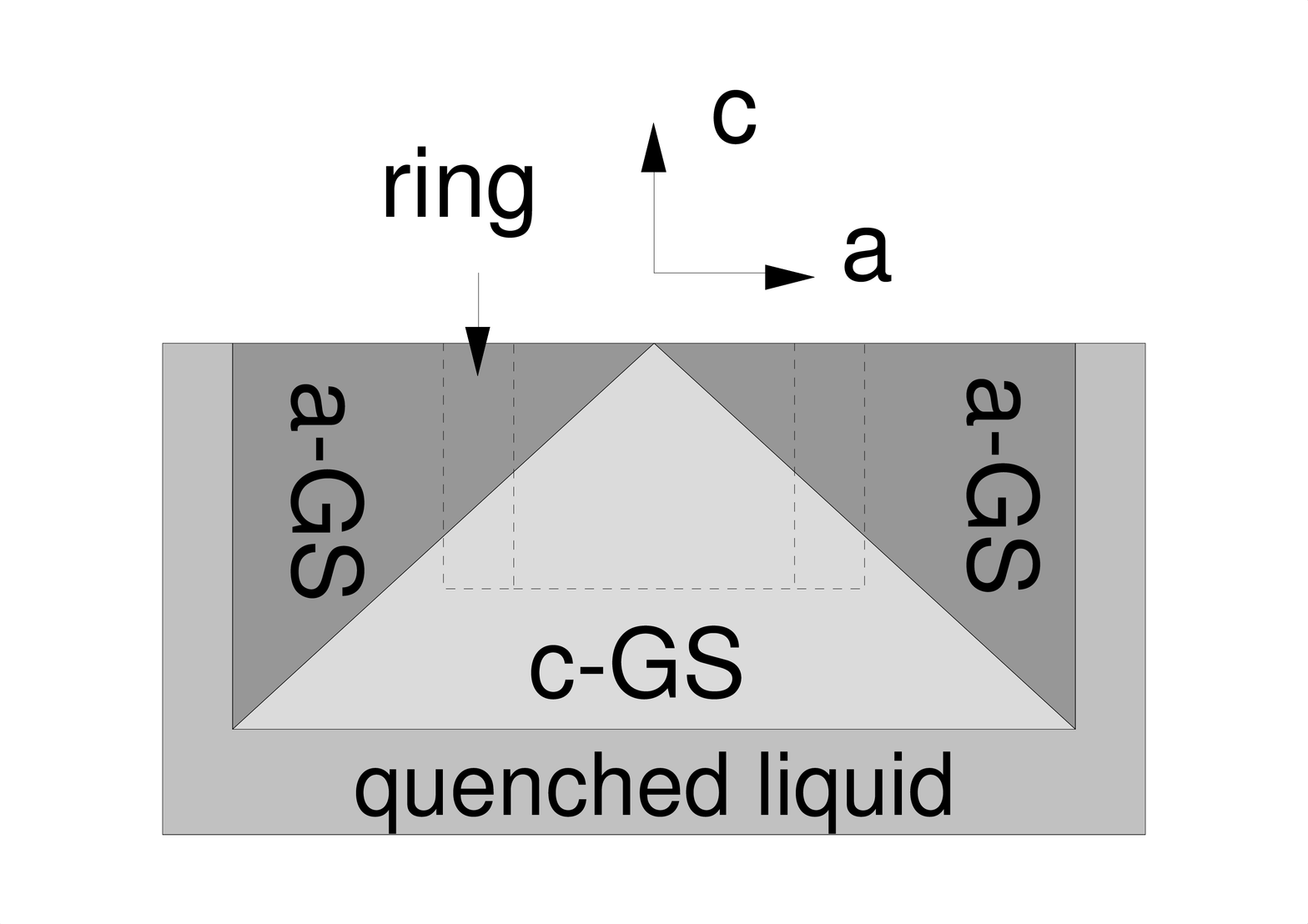}
\end{center}\vspace{\Xmm}
\caption{The central cross-section of the typical MT YBCO crystal.
Note that high rings can cross the boundary between the $a$- and
$c$-sectors grown on different habitus planes, respectively, (100)
and (00$\bar{1}$).} \label{GS}
\end{figure}

\last{In particular, the top-seeded MT crystal grows due to a
motion of five habitus planes, viz., (100), ($\bar{1}00$), (010),
(0$\bar{1}$0) and (00$\bar{1}$) which form, respectively, four
$a$-growth sectors (GS) and the $c$-GS. The latter usually has a
shape of a regular pyramid with a vertex in the seeding point (see
\FIG{GS}). During a growth the interface energy $\Delta\sigma_0$
between the (00$\bar{1}$)-plane and the Y$_2$BaCuO$_5$ (Y211)
in\-clusions appreciably differs from that for the former four
habits~\cite{Diko00}. Since $\Delta\sigma_0$ determines whether
the Y211-particle with a certain size will be expelled or embraced
by the solid-liquid front~\cite{pushing}, the volume fraction
$V_{211}$ of inclusions trapped inside the $c$-GS \emph{always}
appears much smaller than $V_{211}$ in the $a$-GSs~\cite{Diko00}.
Meanwhile, these in\-clusions are well-known~\cite{Y211} to behave
as effective pinning centers. Thus, the \green{intra\-grain}
current density $j_m$ inside the $c$-GS has also to be less than
$j_m$ in $a$-GSs, i.e. near the crystal surface (see \FIG{GS}). At
our growth temperature ($T=988^\circ$C) the height of the $c$-GS
is approximately a half of its base~\cite{Surzhenko02b}. So, high
rings ($2L\geq R_i$) seem to consist}
of the \SC material with different $j_m$. Grinding the ring
bottom, we gradually deleted a part with a worse $j_m$ \green{and,
thereby, made the ring more homogeneous}. One has also to mention
that the $j_m$ values in \TBL{tbl:z0} are averaged over the whole
height $2L$. A genuine ratio of the current density in the $a$-GSs
and that in the $c$-GS may roughly be estimated as a factor of
$2.0\pm 0.2$ \green{which is in a good agreement with this
obtained by magneto\-optical method (see, for example,
\REF{Uspenskaya03})}.

We also tried to estimate the radial in\-homo\-geneities $j_m(r)$
near the top surface of the MT crystals. For this purpose, we
removed a certain portion of \SC material from the outer periphery
of the cut ring CR7 (see \TBL{tbl:z0}) \green{and revealed a vague
tendency for $j_m$ to decrease as $r\rightarrow 0$.}

This brief survey to the growth-induced structure of MT crystals
scarcely seems redundant. At first, it confirms a validity of our
results. It also gives a better understanding of why we used for
further studies only \green{homogeneous rings, i.e. those carved
from the same, $a$-growth sector} (see \TBL{tbl:entire}).
Meantime, it should not eclipse the main, to our opinion,
conclusion of this section: the proposed method well predicts the
``magic position'', $z_0$, where the magnetic flux is free of the
intra\-grain contribution, $B_m$.

\subsection{\label{subsec:entire_remanent} Entire rings in the %
remanent state}

In this section we shall report experimental data for entire rings
which consist of the $a$-twisted sub\-grains. Such sub\-grains are
formed due to mechanical stresses which the YBCO skeleton
undergoes when the growth front en\-gulfes the Y$_2$BaCuO$_5$
inclusions. These stresses result in dislocations which,
unfortunately, have a destructive tendency to amalgamate into the
GBs~\cite{Diko00}. As these stresses accrue during the growth, the
mis-orientation angle between the neighboring sub\-grains (which
are usually elongated along the growth direction, i.e. along
either the $a$- or the $c$-axis) gradually increases with a
distance from the seeding point~\cite{Diko00}. This angle, for
large enough samples, can exceed the critical limit when the GB
turns into the weak link~\cite{Dimos,Chisholm,channel_evidence}.
Having an opportunity to determine whether this limit is
approached or not, we grew few extra-large ($52\times 52\times
25\,mm$) MT crystals and cut entire rings from their various
parts. So, aside from standard ring parameters (i.e. the inner
$R_i$ and the outer $R_o$ radii as well as the full-height, $2L$),
we shall hereafter introduce one more, viz., the distance $X$
between the seed and the center of the ring (see
\TBL{tbl:entire}).
\begin{table}[!b]
\caption{The parameters of entire rings (ER).} \label{tbl:entire}
\begin{ruledtabular}
\begin{tabular*}%
{\hsize}{@{\extracolsep{0ptplus1fil}}ccccccccc@{\extracolsep{0pt}}}
Ring  & $R_o$ & $R_i$ & $2L$ & $X$ & $z_0$ & $j_w$\footnotemark[1]
& $j_m$\footnotemark[1] & $f$\footnotemark[1]\\ \cline{2-6}
\cline{7-8}

No. & && $mm$ &&&\multicolumn{2}{c}{$kA/cm^2$} & \\ \colrule

ER1 & 5.00 & 2.90 & 2.60 & 0  & 2.97 & 13.9$_5$  & \magenta{21.6$_3$} & \magenta{0.64$_5$} \\ 
ER2 & 3.80 & 2.15 & 2.00 & 9  & 2.24 & 16.0$_2$  & \magenta{30.7$_9$} & \magenta{0.52$_0$} \\ 
ER3 & 7.43 & 2.30 & 2.20 & 0  & 3.29 & 19.0$_2$  & \magenta{26.0$_6$} & \magenta{0.73$_0$} \\ 
ER4 & 6.00 & 2.18 & 3.60 & 19 & 3.15 & \ 4.6$_3$ & \magenta{11.3$_6$} & \magenta{0.40$_7$} 

\end{tabular*}
\end{ruledtabular}
\footnotetext[1]{\ The values are calculated from the $B(r=0,z)$
data measured in the remanent state after the long-continued
dwell, $t_{dw}=30\,min$, \magenta{within an assumption of a single
weak link ($N=1$)}.}
\end{table}
When prepared, each ring was carefully inspected with a view to
avoid the cracks which could reduce the effective cross-section
$S=(R_o-R_i)2L$ and influence, thereby, on the current
distribution inside a ring.
\begin{figure}[!tb] \begin{center}
\includegraphics[width=0.85\columnwidth]{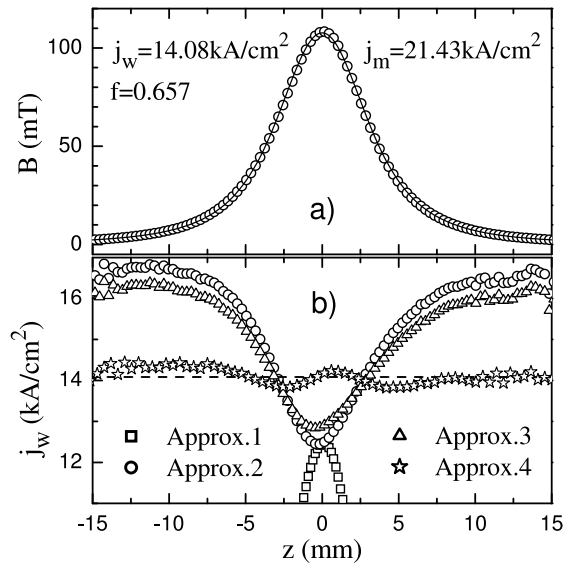}
\end{center}\vspace{\Xmm}
\caption{a) Variation of the remanent flux density $B(r=0,z)$
along the axis of the ring ER1 ($T=77.3\,K$, $t_{dw}=30\,min$) and
the results of the fitting procedure (solid line). b) The
inter\-grain current densities $j_w$ calculated from these data
within various approximations (see details in the text).}
\label{exp_profiles}
\end{figure}

In order to estimate a distribution of the currents flowing in
entire rings at the remanent state, we duplicated the procedure
described in Section~\ref{sec:Samples}. \FIG{exp_profiles}a
presents the profile $B(r=0,z)$ measured along the axis of the
ring ER1 (see \TBL{tbl:entire}). One can, certainly, fit the whole
profile $B(r=0,z)$ by selecting appropriate parameters $j_m$ and
$f$. The best fit is obtained at \magenta{$j_m=21.43\,kA/cm^2$}
and $j_w=14.08\,kA/cm^2$ (solid line in \FIG{exp_profiles}a).
There exists, though, an opportunity to determine the inter- and
intra\-grain currents separately. By substituting the flux
densities $B(r=0,\pm z_0)$ into \EQ{jw}, one obtains quite similar
results, $j_w=14.0$ and $13.9\,kA/cm^2$ for the points $+z_0$ and
$-z_0$, respectively. Estimation of the intra\-grain current
density $j_m$ seems a bit more complicated. Accepting $j_w$ to be
the average value $13.95\,kA/cm^2$, one can use \EQ{BT} to restore
the flux density $B_w(r=0,z)$ at arbitrary distance $z$ (say, in
the ring center, $z=0$). Then, by deducting this value $B_w(0,0)$
from the experimental one $B(0,0)$, one can extract the
intra\-grain flux component, $B_m(0,0)$. Finally, one has to
adjust the parameter $f$ (or $j_m=j_w/f$) in \EQ{BL} so to
approach the same remainder, $B_m(0,0)$. Following this procedure,
we obtained \magenta{$f=0.645$} and
\magenta{$j_m=j_w/f=21.63\,kA/cm^2$}. The latter is in a
reasonable agreement both with the previous estimate
\magenta{($21.43\,kA/cm^2$)} and with the values
\magenta{($\approx 22.7\,kA/cm^2$)} reported for the cut rings
which also were carved from the central part of MT YBCO sample
(see \TBL{tbl:z0}). Similar procedure was applied to the other
rings, the results are included in \TBL{tbl:entire}. Let us,
though, postpone their discussion till the next chapter.

Since any innovation assuredly needs in a careful comparison with
the conventional methods, we paid a special attention to this
task. At first, we considered those techniques (see, for example,
Refs. \onlinecite{Zheng99,Zheng01,Puig01})
\BEA B=\left\{ \begin{array}{ll} I_w/[2R], & \textrm{Approx.\,1}\\
I_w/[2(R^2+z^2)^{1.5}], & \textrm{Approx.\,2} \\
j_w\Phi(L,z,R_o,R_i), & \textrm{Approx.\,3}
\end{array}\right.
\EEA
which neglect the intra\-grain flux component $B_m$. Using each of
the mentioned approximations, we calculated the inter\-grain
current densities $j_w=I_w/[2L(R_o-R_i)]$ from the same
experimental profile $B(r=0,z)$ and plotted these data in
\FIG{exp_profiles}b. It is worth to underline an evident
similarity of the curves obtained within the Approx.~2 and 3 with
those presented in \FIG{example}b. Their deviation from nearly
constant value $j_w$, which corresponds to the case when the
intra\-grain currents are taken into account (Approx.\,4 in
\FIG{exp_profiles}b), make these methods suitable only for rough
estimates even for relatively homogeneous rings like ER1
\magenta{($f\simeq 0.65$)}.
\begin{figure}[!tb] \begin{center}
\includegraphics[angle=-90,width=0.85\columnwidth]{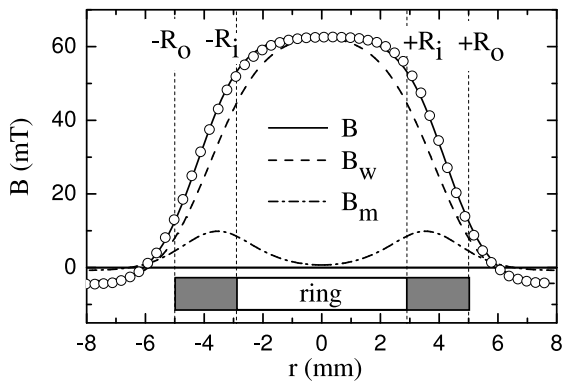}
\end{center}\vspace{\Xmm}
\caption{The radial profile $B^{exp}(r)$ of the remanent flux
density measured at $z=3\,mm$ above the center of the ring ER1 and
its theoretical approximation $B^{calc}(r)$ obtained by adjusting
parameters $j_m$ and $j_w$. The best fit values
$j_m=22.3\,kA/cm^2$ and $j_w=14.05\,kA/cm^2$ ($f=0.63$) were used
to calculate the profiles $B_w(r)$ and $B_m(r)$. Note that since
the distance $z$ is close to $z_0=2.97\,mm$, $B_m(r=0)\approx 0$.}
\label{Bzexp}
\end{figure}

Then, we studied the same ring by the approach which is kin to
that described in \REF{Zheng01}. Briefly, the radial profile
$B(r,z=Const)$ in the remanent state ($T=77.3\,K$,
$t_{dw}=30\,min$) was measured and fitted by adjusting two
parameters, $j_w$ and $j_m$. The obtained data and their fit well
agree one with another (see \FIG{Bzexp}). One has also to mention
that a shape of the inter\-grain flux component $B_w(r)$
noticeably differs from a shape of the experimental curve, $B(r)$.
This example once more demonstrates how important is the flux
correction due to the intra\-grain currents. However, yet more
important, at the moment, is a \magenta{good} agreement between
the best fit parameters, $j_w=14.05\,kA/cm^2$ and
$j_m=22.3\,kA/cm^2$ \gen{($f=0.63$)}, and and their values
estimated by the ``two-point'' method.

Concluding this section, we would like to emphasize the following.
In order to know $j_w$ and $j_m$, one no longer requires to scan
the magnetic flux. Both values may, with a proper accuracy, be
determined from \emph{two} experimental data, e.g., $B(z_0)$ and
$B(0)$. In contrast to any scanning techniques, this ``two-point''
method is quite applicable in a wide range of magnetic fields
and/or temperatures. This also has no evident restrictions for a
time to start the measurements. Realizing these benefits, we used
this approach for further studies.

\subsection{\label{subsec:external-fields}Rings in external magnetic fields}

Heretofore, we acquiesced to the Bean model which presumes that
the critical current density remains independent of the external
magnetic field, $H$. Meantime, $H$ is well-known to reduce $j_w$.
This reduction, in turn, can essentially worsen, say, a
performance of \SC devices constructed from the welded grains. It
is, therefore, vitally important to control the inter\-grain
current $I_w$ which flow through the weak link immersed in strong
magnetic fields. Another task on which we are going to focus an
attention is a penetration of a magnetic flux into the rings. This
process was recently shown~\cite{Pannetier01} to be accompanied by
a highly non-uniform current distribution. Although our approach
describes mostly rings in the full-magnetized state, this allowed
to shed some light onto this phenomenon.

Since a behaviour of the rings enumerated in \TBL{tbl:entire}
\emph{gradually} changed from the most homogeneous case, ER3
\magenta{($f\simeq 0.73$)}, to the most in\-homogeneous one, ER4
\magenta{($f\simeq 0.41$)}, these were the samples which we
selected for an illustration.

\begin{figure}[!tb] \begin{center}
\includegraphics[width=0.85\columnwidth]{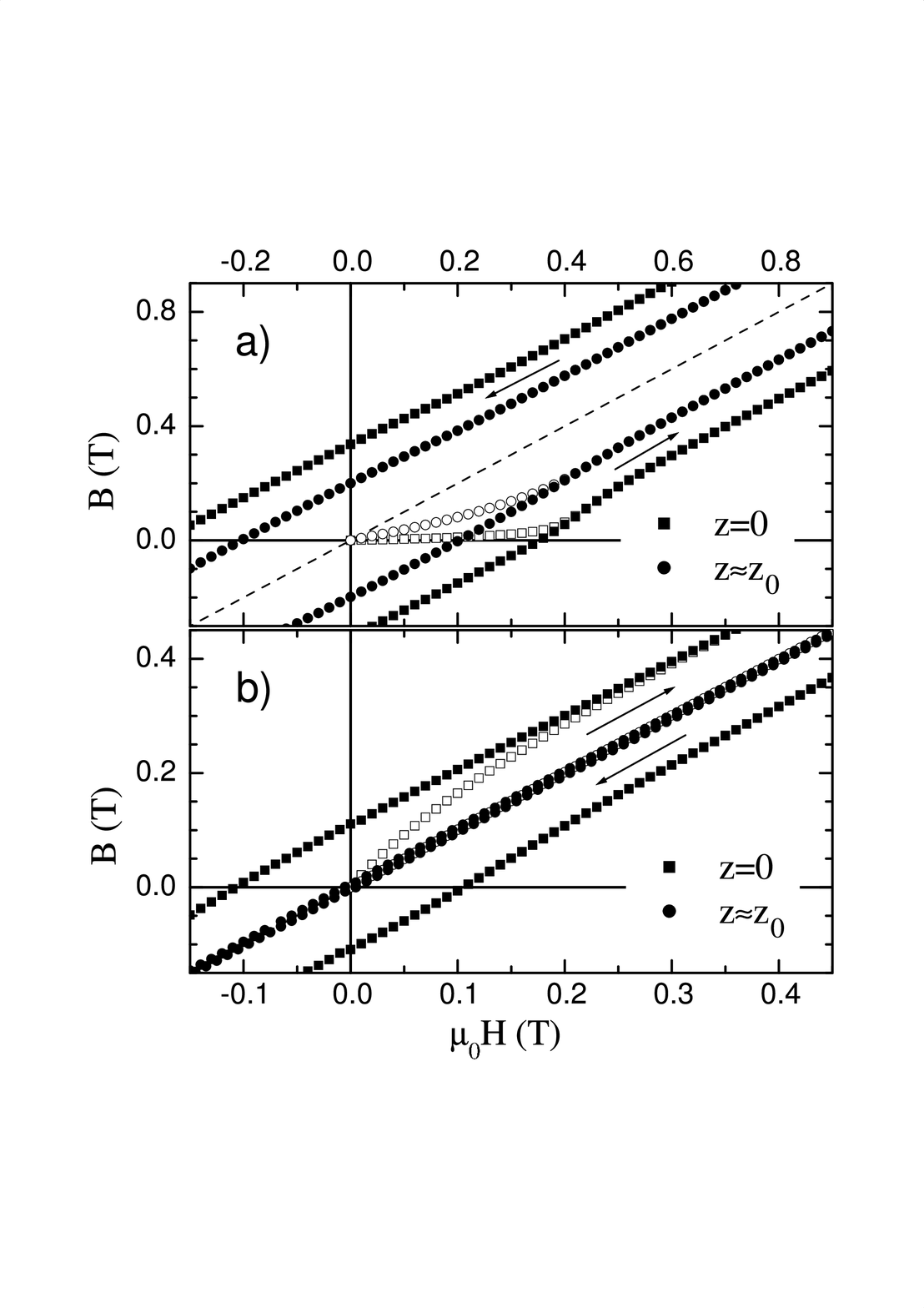}
\end{center}\vspace{\Xmm}
\caption{The $B$ vs $H$ dependencies ($T=77.3\,K$, the sweep rate
$0.5\,T/min$) for the ring ER3 a) without and b) with the radial
slit. The curves were measured at various heights, $z=0$ and
$z_0\approx 3.3\,mm$, above the ring center.} \label{ER3}
\end{figure}

\FIG{ER3} presents the $B$ vs $H$ dependencies obtained at various
heights, $z=0$ and $z=3.3\,mm\approx z_0$, above the center of the
ring ER3 a) before and b) after we introduced therein the radial
slit. Let us primarily discuss the flux penetration into a bore
($z=0$) of the zero-field-cooled (zfc) ring in the non-magnetized,
virgin state (open symbols).

\FIG{ER3}a shows that the entire ring shields ($B=0$) its
bore~\cite{Buckel} by generating the circular current $I_w$ which,
at first, exactly compensates $\mu_0H$. However, $I_w$ is limited
by its critical value. Thus, the field $H_p$, when the currents
are no longer enough for full screening ($B>0$), may be used to
estimate the critical current den\-sity $j_w$.

Penetration of the magnetic flux into a bore ($z=0$) of the cut
rings seems absolutely different. Since there are no circular
currents ($I_w=B_w=0$), the flux density is defined merely by the
intra\-grain flux component, $B_m$. For this reason, the initial
curve $B(z=0)$ in \FIG{ER3}b deviates from zero as soon as
$H>0$. One can also observe that the initial slope $dB/\mu_0dH$
exceeds unity. In other words, the $B(z=0)$ loop in the cut rings
and that in the entire rings have opposite widths $\Delta
B=B-\mu_0H$(\FIG{ER3}). This behavior is reminiscent of the
opposite components, $B_m(z=0)<0$ and $B_w(z=0)>0$, of the
remanent flux presented in \FIG{example}. The $B_m$ vs $z$
dependencies on Figs.\,\ref{example} and \ref{cutrings} also
explain why the loop $B(H)$ in the cut rings shrinks into the
straight line $B=\mu_0H$ as $z$ increases up to $z_0$ (see the
respective curve in \FIG{ER3}b) and, then, expands again with
the negative width, $\Delta B(z>z_0)<0$.

\begin{figure}[!tb] \begin{center}
\includegraphics[angle=-90,width=0.935\columnwidth]{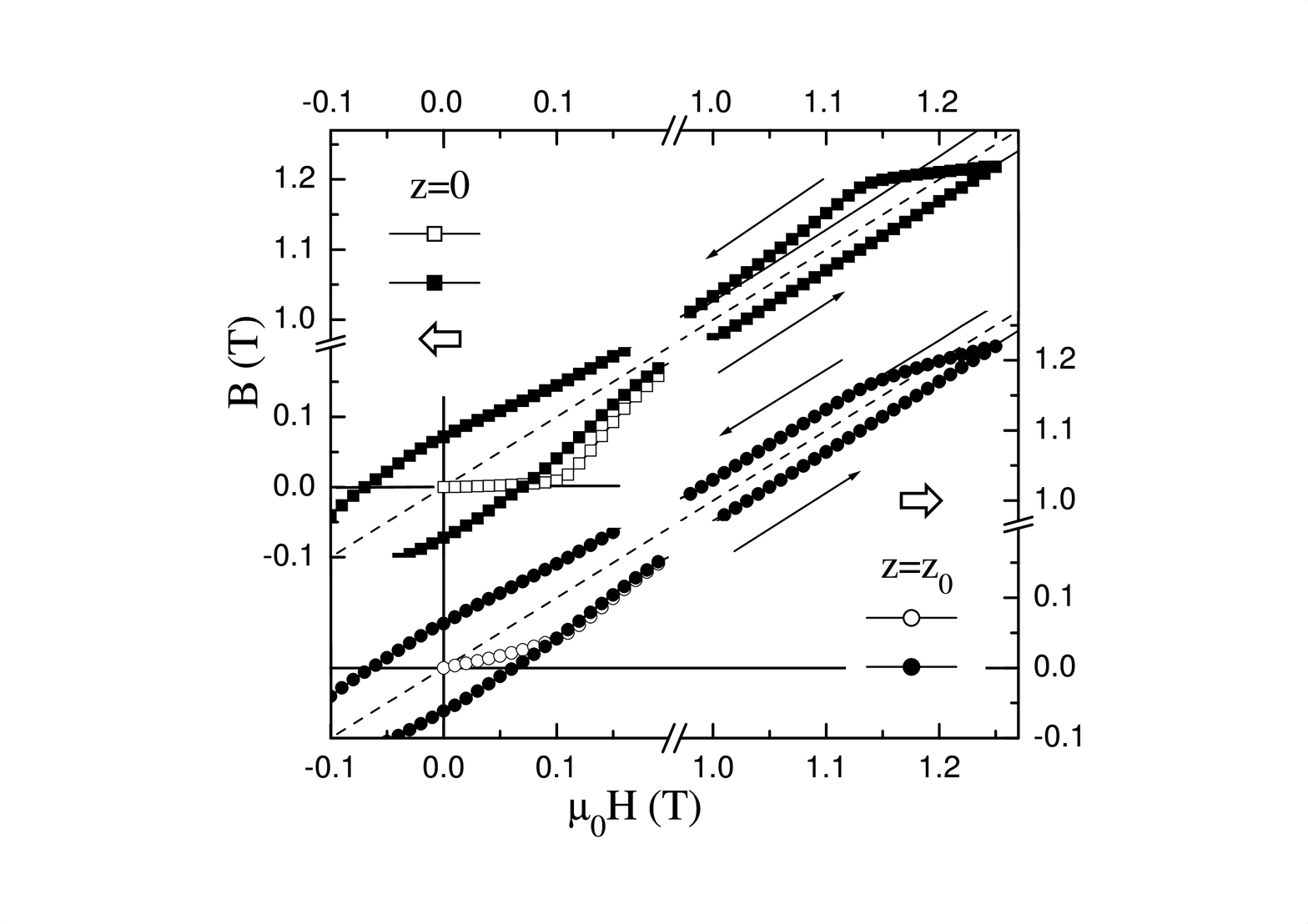}
\end{center}\vspace{\Xmm}
\caption{The $B$ vs $H$ dependencies ($T=77.3\,K$, the sweep rate
$0.5\,T/min$) for the ring ER4. The data are measured at various
heights, $z=0$ and $z_0\approx 3.15\,mm$, above the ring center.
Solid lines (on the right side) correspond to the loops
re-magnetized at higher magnetic field, $\mu_0H=1.5\,T$. Note that
the $B(z=0)$ anomalies during the field penetration (on the
initial branch, $\mu_0H\approx 0.12\,T$, as well as on the
descending field branch, near the field reversal) are absent at
the height $z=z_0$.} \label{ER4}
\end{figure}

Extremely interesting anomalies (see \FIG{ER4}) accompany the flux
penetration into a bore ($z=0$) of the ring ER4, where the weakest
link is really weak \magenta{($f\simeq 0.41$)} as compared with
the rest of the ring material. In particular, the curve of the
magnetic flux density $B(z=0)$ at the initial field branch (open
squares) approaches the full-magnetized loop (solid squares) from
outside, i.e. there exist the field range
($0.08\,T\leq\mu_0H\alt0.5\,T$) wherein the partially magnetized
ring more effectively screens its bore than that in the
full-magnetized state. Similar anomalies are also observed near
the field reversal. \FIG{ER4} shows these effects at
$\mu_0H=1.25\,T$, solid lines on the right side of \FIG{ER4}
correspond to the loop reversed at a bit higher magnetic field,
$\mu_0H=1.5\,T$. Since these anomalies almost disappear at the
height $z=z_0$ (circles), one can preliminary conclude that these
are induced by changes in a distribution of the
\emph{intra\-grain} currents which correspond to the concentric,
counter\-rotating current loops mentioned in \REF{Pannetier01}.

\begin{figure}[!tb] \begin{center}
\includegraphics[angle=-90,width=0.9\columnwidth]{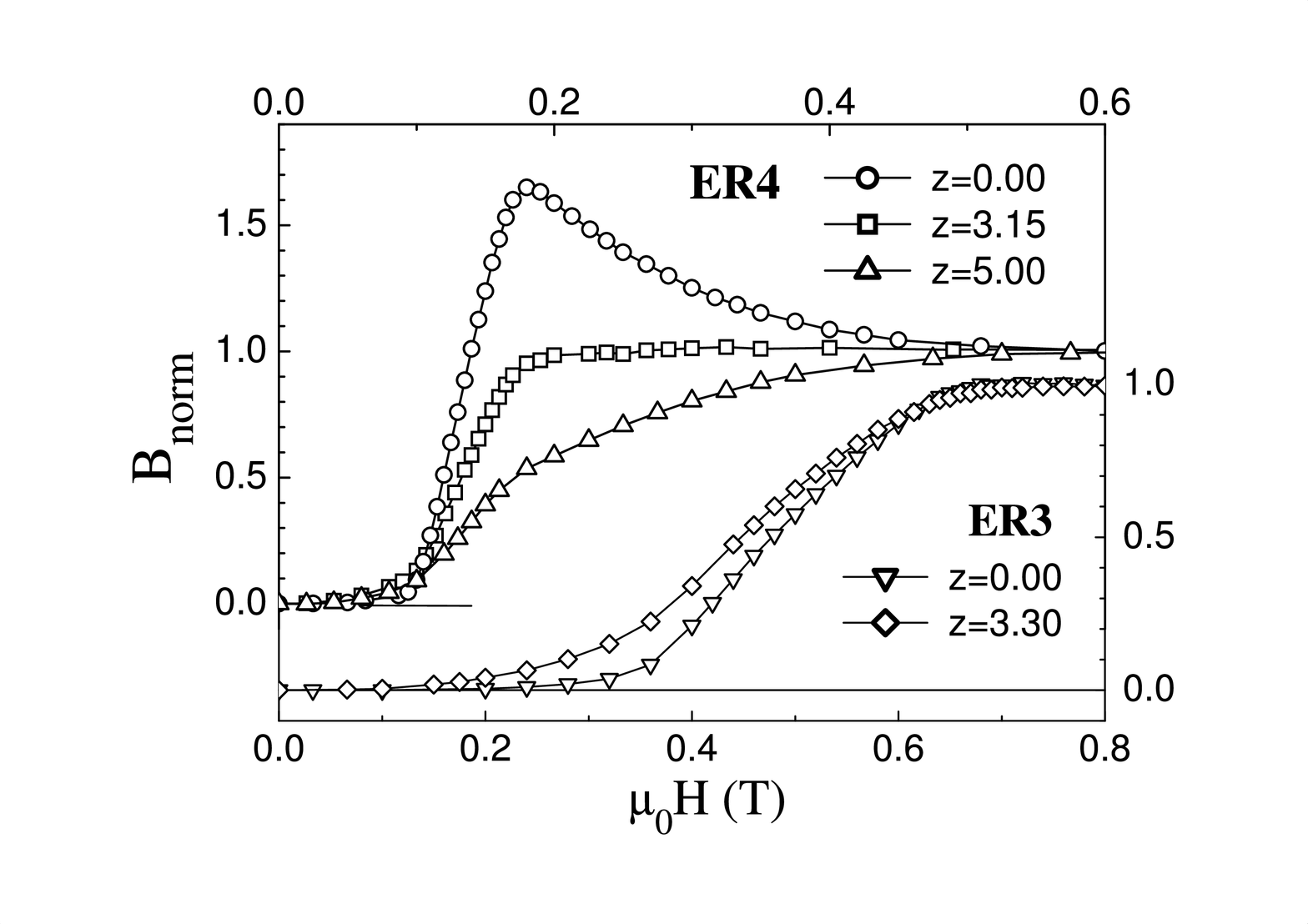}
\end{center}\vspace{\Xmm}
\caption{The remanent flux density $B$ vs the magnetic field
$\mu_0H$ at which the zfc-rings, ER3 (right and bottom axes) and
ER4 (left and top axes), were magnetized ($T=77.3\,K$). The curves
correspond to various distances, $z=0$, $z=z_0$ and $z>z_0$ (given
in $mm$), where the flux density was registered after the short
dwell, $t_{dw}=10\,s$. Each curve is normalized to the
full-penetration value obtained at $\mu_0H=2\,T$.} \label{penetr}
\end{figure}

To test this suggestion, we realized the following experiments.
The samples were zfc-cooled down to $T=77.3\,K$ and magnetized at
this temperature in a certain field, $H$. When a field was
switched off, the remanent flux density was registered (after
$t_{dw}=10\,s$) at various heights above the ring center, $z=0$,
$z_0$ and $z>z_0$. We repeated this procedure at gradually
increasing $H$ until the remanent flux density $B$ approached its
saturation. The obtained $B$ vs $\mu_0H$ dependencies (normalized
to the $B$ values at $\mu_0H=2\,T$ which well exceeds the
full-pene\-tra\-tion field) are presented in \FIG{penetr}. One can
see that the flux density in the center of the weakly coupled ring
ER4 (open circles) does exhibit a well-discernible maximum at
$\mu_0H=0.18\,T$. Similar maximum was already observed in
poly\-crystalline YBCO rings by Darhmaoui and
Jung\cite{Darhmaoui96}. However, the concept of the Josephson
vortex flow, which they invoked to unravel this phenomenon, can
scarcely explain why the maximum totally disappears at larger
distances, $z=z_0$ and $z>z_0$ (respectively, squares and up
triangles in \FIG{penetr}).

To our opinion, both the maximum $B(z=0)$ in \FIG{penetr} and the
$B(z=0)$ anomalies in \FIG{ER4} are of the same origin. Their
reason becomes nearly obvious owing to the normalization trick
used in \FIG{penetr}. These phenomena occur since the
full-pene\-tra\-tion field $\mu_0H_p\approx 0.5\,T$ (which is
necessary to induce the whole currents) in the weakly coupled ring
ER4 is much larger than the field $\mu_0H_w\approx 0.2\,T$ wherein
the inter\-grain current reaches its critical limit,
$j_w2L(R_o-R_i)$ (squares in \FIG{ER4}). So, a further growth of
magnetic field ($H_w\leq H\alt H_p$) can generate merely the
intra\-grain currents. Since the intra\-grain flux component $B_m$
reverses its sign at the height $z=z_0$ (see, for example,
Figs.\,\ref{example}, \ref{cutrings} or \ref{ER3}b), the increase
of its amplitude has to result in an additional increment of the
net flux density $B$ measured at $z>z_0$ and, what we are going to
prove, in its reduction when $z<z_0$. In other words, we believe
that these anomalies become visible owing to de\-magnetizing
effects. One more argument in favor of this scenario is that the
ratio $H_w/H_p\simeq 0.4$ does appear very close to the parameter
\magenta{$f\simeq 0.41$}. The same scenario is quite relevant for
explication of the flux anomaly after the field reversal.

Generally, a certain decrease of the remanent flux could also be
explained by the sample heating due to viscous forces that exert
on moving flux lines. Yet, to attain the difference of $65\%$
shown in \FIG{penetr}, one usually needs in the pulsed fields when
the sweep rate is in a few orders of magnitude
faster~\cite{Surzhenko01} than that ($0.5\,T/min$) used in our
experiments. Another argument, which completely rejects a
hypothesis of the sample heating, is an absence of \emph{any}
$B(z=0)$ anomalies for more homogeneous ring ER3
\magenta{($f\simeq 0.73$)} which was measured at the same
experimental conditions (\FIG{penetr}). This absence may readily
be explained by the following reasons. At first, the critical
value of the inter\-grain current $I_w\sim f$ linearly grows with
$f$ and so is the flux component $B_w$ which $I_w$ produces. Since
the intra\-grain current $I_m\sim (1-f^2)$ has an opposite
tendency, the ratio $B_m/B_w$ rapidly decreases as $f\rightarrow
1$. Finally, the larger is $f$, the closer become the fields $H_w$
and $H_p$ where, respectively, the the inter- and the intra\-grain
currents are saturated. Hence, the larger half ($\sim f$) of the
intra\-grain flux component $B_m$ appears \emph{simultaneously}
with the inter\-grain one ($H<H_w$) and only a small portion,
$\sim (1-f)$, does in the range $H_w\leq H\alt H_p$ where the
$B_m$ changes are not compensated by the $B_w$ growth.

Thus, the closer is $f$ to unity, the less pronounced have to be
the $B(z=0)$ anomalies during the flux penetration. The studies of
other rings (enumerated in \TBL{tbl:entire}) confirm this
statement. In particular, a difference between the maximum of the
remanent flux $B(z=0)$ and its saturation value for the ring ER2
was registered to be $3.6\%$ \magenta{($H_w/H_p=0.6\approx
0.52=f$)}. For the rings ER1 and ER3 with \magenta{$f\simeq 0.65$}
and \magenta{$f\simeq 0.73$}, respectively, this value did not
exceed the experimental error ($\approx 0.3\%$). On the other
hand, one can conclude, be the conclusion ever so amazing, that
the remanent flux density $B(z<z_0)$ of weakly coupled rings
($f\rightarrow 0$) may totally vanish or even become negative
(note, for example, a negative remanence of $\approx -0.1\,T$ in
the center of the cut ring ER3 shown in \FIG{ER3}b) as the
magnetizing field ascends.

Let us yet come back to an estimation of the inter\-grain and the
intra\-grain current densities in external magnetic fields. As
mentioned above, our approach requires removing the constant
background $\mu_0H$ from the experimental data, $B(z=0)$ and
$B(z=z_0)$, in the full-magnetized state (full points in
Figs.\,\ref{ER3}a and \ref{ER4}). This task may conveniently be
resolved by calculating the half-difference between the
descending, $B_\downarrow$, and the ascending, $B_\uparrow$, field
branches
\BE \Delta B(H)=[B_\downarrow(H)-B_\uparrow(H)]/2.
\label{loop_width} \EE
\begin{figure}[!tb] \begin{center}
\includegraphics[width=0.85\columnwidth]{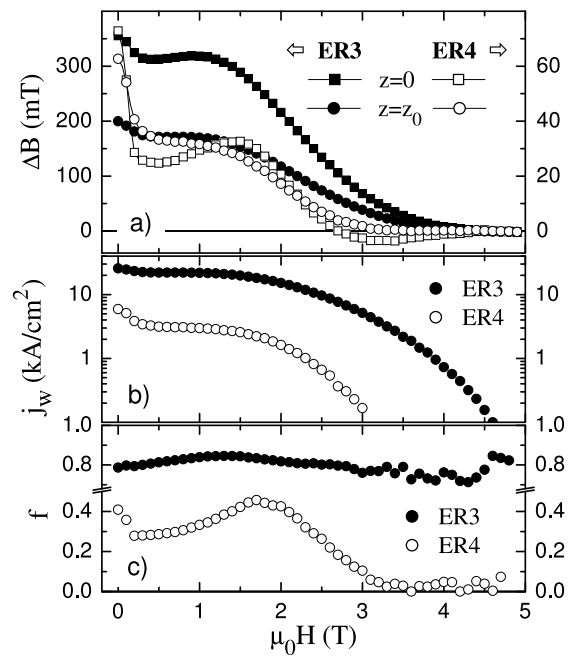}
\end{center}\vspace{\Xmm}
\caption{a) The half-width $\Delta B(H)$ of the flux density loops
shown in Figs.\,\ref{ER3}a and \ref{ER4}. The field dependencies
b) $j_w(H)$ and c) $f(H)$ are calculated from the $\Delta B(H)$
data.} \label{j}
\end{figure}
Then, by applying the ``two-point'' method to each pair of the
points, $\Delta B(H)|_{z=0}$ and $\Delta B(H)|_{z=z_0}$ presented
in \FIG{j}a, we restored the sought $j_w(H)$ and $j_m(H)$
dependencies as well as their ratio $f(H)$ which also was found to
depend on $H$. Moreover, the weaker is the link, the more
sensitive this appeared to external magnetic fields (see
\FIG{j}c). In particular, in contrast to the sample ER3 which
values $f$ cluster around 0.8, the curve $f(H)$ in the ring ER4
exhibits two well-defined maxima. Their origin may be attributed
to the strongly coupled channels in GBs with a low
mis\-orientation angle~\cite{Dimos,Chisholm,channel_evidence}.
\gen{It is suggested ~\cite{Dimos,Chisholm} that} small magnetic
fields \emph{partially} de\-couple a weak link, i.e. there exists
a secondary, non-weak-linked component of the inter\-grain
conduction. Such channels of relatively un\-disturbed crystal
lattice (i.e. micro\-bridges of the intrinsic, intra\-granular
material which occupy only a small area of the GB surface) were
directly confirmed by the transmission electron microscopy of the
$10^\circ$-bi\-crystal YBCO film~\cite{channel_evidence}.
Following this multi\-filamentary model, one can identify the
saddle point between two mentioned maxima as the field which
de\-couples the weakest link, but still does not influence onto
the other weak links restricting the intra\-grain current $I_m$
given by \EQ{Im}. So, a further increase of field reduces mostly
the intra\-grain current density, $j_m(H)$, and, hence, results in
the second peak of the $f(H)$ dependence. More interesting is that
the same sequence seems valid for high fields $H_{irr}$ which
totally de\-couples weak links by breaking their strongly coupled
channels. \FIG{j}c shows that this de\-coupling first happens with
the weakest link $\mu_0H_{irr}\approx 3.3\,T$ and, then, with the
rest of weak links ($\approx 4.6\,T$) in the ring ER4. In other
words, within the range $3.3\,T\alt \mu_0 H \alt 4.6\,T$ there is
already no circular current ($f=0$), but the intra\-grain
currents, which result in $B_m<0$, still continue to flow. It is
the reason why the half-width $\Delta B(H)|_{z=0}$ (open squares
in \FIG{j}a) alters its sign. To our knowledge, only this type of
measurements opens up an opportunity to see this high-field
phenomenon.

Finalizing this section, one has to mention some discrepancies
between the parameters $f(H=0)$ extracted from the magneti\-zation
cycle data (\FIG{j}c) and those which were registered after a long
dwell (see \TBL{tbl:entire}). Comparing these values, we took
cognizance of a weak tendency for $f$ to decrease with a time.
This question, though, is worthy of a separate discussion.

\subsection{\label{subsec:dissipation}Dissipation of the magnetic flux}

It is well known that the thermo\-activated vortices ($T>0$) can
leave the pinning centers under the Lorentz force~\cite{creep}
\BE F_L=V \frac{B}{\mu_0}\frac{dB}{dr}. \label{FL} \EE
Such motion of the vortex lines is equivalent to a resistance to a
current and responsible for the power dissipation. For this
reason, we were not surprised that the absolute values of the
inter- and the intra\-grain current densities were found to
decrease during a dwell. Meanwhile, their ratio $f$ was also
suspected to depend on a time. If this tendency does exist, the
flux de\-pinning processes may noticeably falsify a genuine
ability $f$ of weak links to transmit the current. It is therefore
very important to know to which extent can we rely on the previous
data obtained by the scanning Hall-sensor
magneto\-metry%
~\cite{Philip98,Zheng99,Delamare00,Zheng01,Prikhna01,Harnois01,%
Puig01,Yoshioka02} which \emph{a-priori} requires a long-continued
dwell.

\begin{figure}[!t] \begin{center}
\includegraphics[angle=-90,width=0.99\columnwidth]{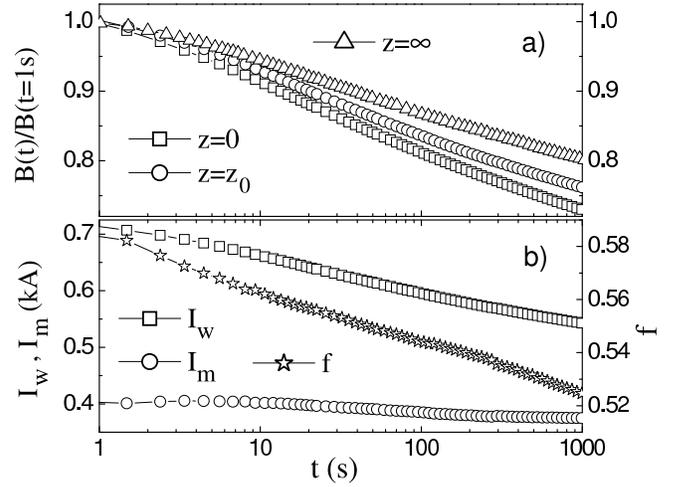}
\end{center}\vspace{\Xmm}
\caption{a) Dissipation of the remanent flux density $B(t)$
measured at various heights $z$ above the center of the ring ER2
($T=77.3\,K$). The curve $z=\infty$ corresponds to the time
changes of the remanent magnetic moment, $m$. b) The inter- and
the intra\-grain currents, $I_w$ and $I_m$, as well as the
parameter $f=j_w/j_m$ which are restored from these data.}
\label{diss_ER2}
\end{figure}

At first, we more carefully explored the inter- and the
intra\-grain flux losses and their influence on the current
distribution. \FIG{diss_ER2}a shows the time dependencies of the
remanent flux density measured at various heights, $z=0,\ z_0$ and
$\infty$ above the center of the ring ER2. These curves
(normalized, for convenience, to their values $B_0=B(t=1\,s)$)
neatly indicate that the relative dissipation rate
\BE S=-\frac{1}{B_0}\frac{d B(t)}{d\log{t}} \nonumber \EE
gradually decreases with a distance $z$ from the sample. This
decrease does prove that the inter\-grain current $I_w$ is
dissipated faster than the intra\-grain one, $I_m$. Using the
``two-point'' procedure, we confirmed this qualitative conclusion
by the numerical values $I_w$ and $I_m$ (\FIG{diss_ER2}b). The
inter\-grain current decay $S$ was determined to be 8.2\% per the
time decade, whereas $S$ for the intra\-grain current did not
exceed 2\%. 
Consequently, \gen{each sequent} time decade reduced the parameter
$f$ on 0.021.

There may be several reasons which explain this phenomenon. The
most evident one is that the intra\-grain current loop does not
cross the weakest link where the largest losses are expected. One
can also attribute this effect to the Lorentz force (\ref{FL})
which, in accordance to the radial distribution $B(r)$ (see
\FIG{profiles}a), more effectively stimulates a motion of the flux
vortices trapped near the inner periphery ($R_c<r<R_i$) of the
ring. This question will be the subject of next studies.

\begin{figure}[!b] \begin{center}
\includegraphics[angle=-90,width=0.9\columnwidth]{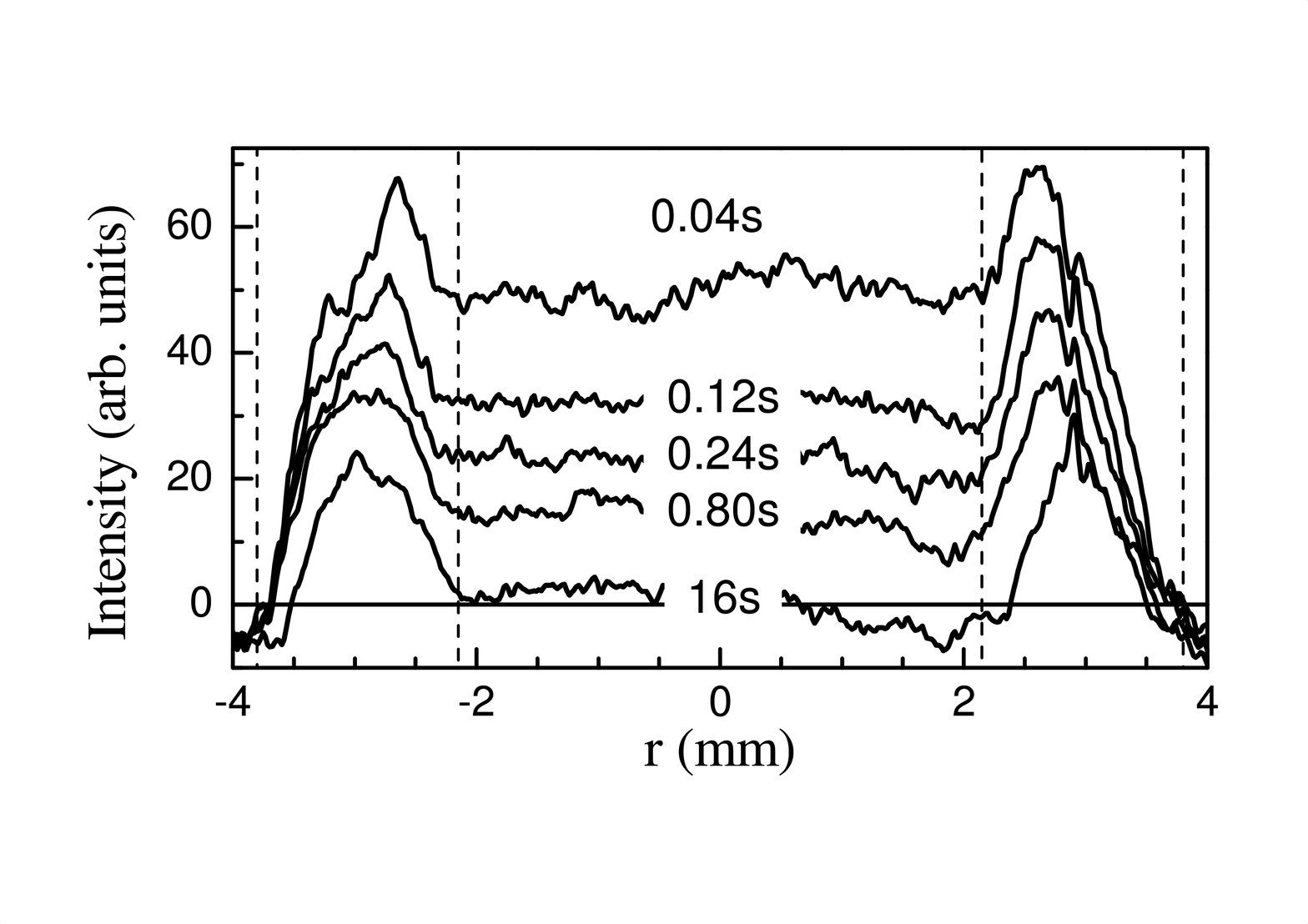} %
\end{center} \vspace{\Xmm}
\caption{Radial profiles of the remanent flux density $B(r,z)$ on
the surface $z=\pm L$ of the ring ER2 at various dwell times (in
seconds). The data were obtained by the magneto-optical imaging
technique ($T=81\,K$).} \label{optics}
\end{figure}
Whatever are the genuine reasons, these do result in the spatial
re-distribution ($f\neq Const$) of the currents during a dwell.
This re-distribution passes the faster, the faster is the flux
creep itself. One can conclude that at low temperatures and small
magnetic fields, slow experimental methods still may give a
reasonable approximation for the parameter $f$ describing the
current limiting properties of the weak link. Otherwise, the
long-continued dwell is absolutely in\-admissible and so,
unfortunately, is the \emph{scanning} Hall-sensor method, which
provides another way to discern between the inter- and the
intra\-grain currents~\cite{Zheng01}. In order to visualize how
appallingly large may become an error just after a few seconds of
dwell, we investigated the remanent flux creep in same ring ER2 at
higher temperature, $T=81\,K$. In view of extremely sharp decay,
we were caused to use the magneto-optical image
technique~\cite{Uspenskaya97} which allowed to register the flux
changes much faster than other methods. So, the ring was cooled
down to $81\,K$, iso\-thermally magnetized and, then, a magnetic
field was rapidly (during a delay $0.04\,s$ between neighboring
image frames) switched off. \FIG{optics} clearly shows that the
flux inside the ring bore $|r|<2.15\,mm$ (dashed lines), i.e. that
which is induced by the inter\-grain (circular) current $I_w$,
rapidly decreases \gen{with a time and so does the parameter $f$}.

\section{\label{sec:Summary} Summary}

We explored the inter- and the intra\-grain currents flowing in
bulk MT YBCO samples containing weak links.

\begin{figure}[!bt] \begin{center}
\includegraphics[angle=-90,width=0.85\columnwidth]{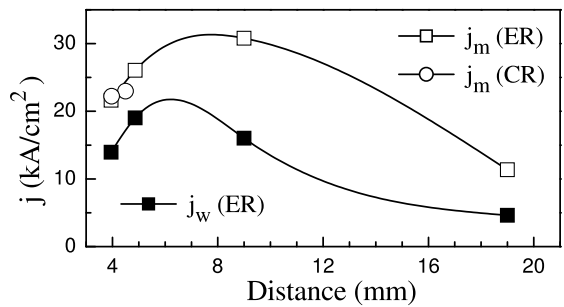} %
\end{center} \vspace{\Xmm}
\caption{The inter- and the intra\-grain current densities inside
the MT YBCO crystal vs the distance from the seeding point.
\gen{For those rings which are carved from the central part of the
crystal, the distance is taken equal to their average radius
$(R_o+R_i)/2$}. The data are compiled from measurements of both
the entire rings (ER) and the cut ones (CR).}
\label{inhomo}\vspace{\Xmm}
\end{figure}

To perform these studies, we developed a simple, reliable and
non-destructive method which allows to discern between these types
of current. In particular, we showed that, if a sample has a shape
of a ring with finite sizes, the intra\-grain current (because of
de\-magne\-tizing effects) gives an alternating-sign contribution
$B_m$ to the net flux density $B$, viz., $B_m$ is negative near
the ring center and positive at large distances. We also
considered the points $z=\pm z_0$, wherein the ring axis ($r=0$)
intersects the border ($B_m=0$) between these areas, and
demonstrated that the distance $z_0$ appears almost independent of
the current distribution. Hence, positioning the Hall sensor at
the height $\pm z_0$ above (or below) the ring center, one can
measure the ``pure'' inter\-grain flux component, $B_w$, and
estimate the inter\-grain (circular) current which produces $B_w$.
This, the so-called ``one-point'' method seems very promising for
characterization of a quality of \SC welds. In this case, the
numerical calculations of the distance $z_0$ may optionally be
replaced by its direct measurement preceding to the welding
procedure. The same idea may, in future, be extended for less
exotic sample shapes, for example, for the finite-size slabs.

By measuring the flux density $B(r=0,z)$ in one more point $z\neq
z_0$, one can also restore the intra\-grain current, $I_m$. We
showed that it is the current which is responsible for very
unusual phenomena registered near the center $(r=0,z=0)$ of
weakly coupled rings ($f<0.5$). In particular, we observed (see
Figs.\,\ref{ER4} and \ref{penetr}) that, being in partially
penetrated state, such rings screen their bores more effectively
than those after the full-flux-penetration. Similar situation was
registered after the field reversal at relatively high fields
(\FIG{ER4}). We also revealed, to the first blush, absolutely
irregular situation when, owing to a destruction of the strongly
coupled channels~\cite{Dimos,Chisholm,channel_evidence} in the
weakest link, magnetic fields $H\geq H_{irr}$ result in a negative
screening of the ring bore (\FIG{j}a). The dependence of the flux
dissipation rate $S$ on a distance $z$ from the ring
(\FIG{diss_ER2}a) could also be scarcely explicable unless the
intra\-grain current $I_m$ is taken into account. These effects
are merely the first items in a large list of somewhat ``strange''
effects (both already published and still waiting for publication)
in the finite-size \SC rings and/or hollow cylinders. We believe
that, by presenting the method which allows to estimate \gen{$I_w$
and} $I_m$, this work opens new perspectives to reveal and to
explain other items.

Concluding the paper, we would like to touch on the \SC welding
again. Unless the welding techniques give totally reproducible
quality of joints, one has to minimize their number by welding the
bulk MT crystals which are as large as possible. Their sizes,
though, are severely restricted because of a degradation of the
\SC properties from the seeding point toward the crystal rims (see
\FIG{inhomo}). There exist, therefore, an optimum \gen{welding}
size, i.e. the size at which the rims of welded parts may transmit
the same density of the current as that which can flow across
\gen{their joint}. Since, at present, the best joints are reported
to carry the currents with a density of $\approx 10\,kA/cm^2$, we
believe that the welding technology should be developed mostly for
joining of relatively large, $25\times 25\,mm$ MT YBCO blocks.

\begin{acknowledgments} This work was supported by the German
BMBF under the Project 13N6854A3, the Russian Foundation for Basic
Research (Project 02--02--17062) and \gen{INTAS (Project
02--2282)}. One of the authors (A.B.S.) would like to thank
R.~Hiergeist for useful discussions.\end{acknowledgments}





\begin{thebibliography}{100}

\bibitem{Leiderer88}P.~Leiderer and R.~Feile, \ZPB {\bf 70}, 141
(1988).

\bibitem{Mohamed91} M.~A.-K.~Mohamed and J.~Jung, \PRB {\bf 44},
4512 (1991.)

\bibitem{Jung93}J.~Jung, I.~Isaac and M.~A.-K.~Mohamed, \PRB {\bf 48},
7526 (1993). 

\bibitem{Darhmaoui96}H.~Darhmaoui and J.~Jung, \PRB {\bf 53},
14621 (1996).

\bibitem{Jin88} S.~Jin, T.~H.~Tiefel, R.~C.~Sherwood, M.~E.~Davis,
R.~B.~van Dover, G.~W.~Kammlott, R.~Fastnacht and
H.~D.~Keith, \APL {\bf 52}, 2074 (1988).~

\bibitem{Salama92} M.~J.~Sturm, Z.~A.~Chaudury
and S.~A.~Akbar, \ML {\bf 12}, 316 (1991); K.~Salama and V.
Selvamanickam, \JAP {\bf 60}(7), 898 (1992).

\bibitem{Shi95} D.~Shi, \APL {\bf 66}, 2573 (1995).

\bibitem{Philip98} Ph.~Vanderbemden, A.~D.~Bradley,
R.~A.~Doyle, W.~Lo, D.~M.~Astill, D.~A.~Cardwell and A.~M.
Campbell, \PC {\bf 302}, 257 (1998).

\bibitem{Zheng99} H.~Zheng, M.~Jiang, R.~Nikolova, U.~Welp,
A.~P.~Paulikas, Yi Huang, G.~W.~Crabtree, B.~W.~Veal and H.~Klaus,
\PC {\bf 322}, 1 (1999).

\bibitem{Zheng01} H.~Zheng, H.~Claus, L.~Chen,
A.~P.~Paulikas, B.~W.~Veal, B.~Olsson, A.~Koshelev, J.~Hull and G.
W.~Crabtree, \PC {\bf 350}, 17 (2001).

\bibitem{Prikhna01} T.~Prikhna, W.~Gawalek, V.~Moshchil,
A.~Surzhenko, A.~Kordyuk, D.~Litzkendorf, S.~Dub, V.~Melnikov, A.
Plyushchay, N.~Sergienko, A.~Koval, S.~Bokoch and T.~Habisreuther,
\PC {\bf 354}, 333 (2001).

\bibitem{Delamare00} M.~P.~Delamare, H.~Walter, B.~Bringmann,
A.~Leenders and H.~C.~Freyhardt, \PC {\bf 329}, 160 (2000).

\bibitem{Harnois01} C.~Harnois, G.~Desgardin and X.~Chaud,
\SUST {\bf 14}, 708 (2001).

\bibitem{Puig01} T.~Puig, P.~Rodriguez Jr., A.~E.~Carillo,
X.~Obradors, H.~Zheng, U.~Welp, L.~Chen, H.~Claus, B.~W.~Veal and
G.~W.~Crabtree, \PC {\bf 363}, 75 (2001).

\bibitem{Yoshioka02} J.~Yoshioka, K.~Iida, T.~Negichi,
N.~Sakai, K.~Noto and M.~Murakami, \SUST {\bf 15}, 712 (2002).

\bibitem{Noudem01} J.~G.~Noudem, E.~S.
Reddy, M.~Tarka, M.~Noe and G.~J.~Schmitz, \SUST {\bf 14}, 363
(2001).

\bibitem{Walter01} H.~Walter, Ch.~Jooss, F.~Sandiumenge,
B.~Bringmann, M.~P.~Delamare, A.~Leenders and H.~C.~Freyhardt,
\EPL {\bf 55}(1), 100 (2001).

\bibitem{Claus01} H.~Claus, U.~Welp, H.~Zheng, L.~Chen,
A.~P.~Paulikas, B.~W.~Veal, K.~E.~Gray and G.~W.~Crabtree,
\PRB{\bf 64}, 144507 (2001).

\bibitem{Kord01} A.~A.~Kordyuk, V.~V.~Nemoshkalenko,
A.~I.~Plyush\-chay, T.~A.~Prikhna and W.~Gawalek, \SUST {\bf 14},
L41 (2001).

\bibitem{Kambara02} M.~Kambara, N.~Hari Babu, D.~A.~Cardwell and
A.~M.~Campbell, \PC {\bf 372--376}, 1155 (2002).

\bibitem{Murakami00} M.~Murakami, \SUST {\bf 13}, 448 (2000).

\bibitem{Pannetier01} M.~Pannetier, F.~C.~Klaassen,
R.~J.~Wijngaarden, M.~Welling, K.~Heeck, J.~M.~Huijbregtse, B.~Dam
and R.~Griessen, \PRB{\bf 64}, 144505 (2001).

\bibitem{Surzhenko02a} A.~B.~Surzhenko, M.~Zeisberger,
T.~Habisreuther, D.~Litz\-kendorf and W.~Gawalek, \SUST {\bf 15},
1353 (2002).

\bibitem{Dewhurst98} C.~D.~Dewhurst, Wai Lo, Y.~H.~Shi and D.~A.~Cardwell,
\MSE B{\bf 53}, 169 (1998).

\bibitem{Doris} D.~Litzkendorf, T.~Habisreuther, M.~Wu, T.~Stra\ss er,
M.~Zeisberger, W.~Gawalek, M.~Helbig and P.~G\"ornert, \MSE B{\bf
53}, 75 (1998).

\bibitem{Klupsch} Th.~Klupsch, Th.~Stra\ss er, T.~Habisreuther, W.
Gawa\-lek, S.~Gruss, H.~May, R.~Palka, and F.~J.~Mora Serrano,
\JAP {\bf 82}, 3035 (1997).

\bibitem{Uspenskaya97} L.~S.~Uspenskaya, V.~K.~Vlasko-Vlasov,
V.~I.~Niki\-ten\-ko, and T.~H.~Johansen, \PRB {\bf 56}, 11979
(1997).

\green{
\bibitem{Uspenskaya03} L.~S.~Uspenskaya, I.~G.~Naumenko, G.~A.~Emelchenko,
Yu.~B. Bu\-go\-slav\-skii, S.~A.~Zver'kov, E.~B.~Yakimov,
D.~Litzkendorf, W.~Gawalek and A.~D.~Caplin, \PC {\bf 390}, 127
(2003).}

\bibitem{Diko00} P.~Diko, \SUST {\bf 13}, 1202 (2000).

\bibitem{Surzhenko02b} A.~B.~Surzhenko, S.~Schauroth,
M.~Zeisberger, T.~Habisreuther, D.~Litzkendorf and W.~Gawalek, \PC
{\bf 372-376}, 1212 (2002).

\bibitem{pushing} C.~Kim, H.~G.~Lee, K.~B.~Kim and G.~W.~Hong,
\JMR {\bf 10} 1605 (1995). 

\bibitem{Y211} Whether the vortex pinning by the
Y211-inclusions is due to the pure interfacial mechanism (see
M.~Murakami (ed.), \emph{Melt processed high-temperature
superconductors} (World Scientific, Singapore, 1992)) or to the
YBCO dislocations which surround the embraced Y211-inclusions
(K.~Salama and D.~F. Lee, \SUST {\bf 7}, 177 (1994)) is still
ambiguous.

\bibitem{Dimos} D.~Dimos, P.~Chaudhari,
J.~Mannhart and F.~K.~LeGoues, \PRL {\bf 61}, 219 (1988);
D.~Dimos, P.~Chaudhari and J.~Mannhart, \PRB {\bf 41}, 4038
(1990).

\bibitem{Chisholm} M.~F.~Chisholm and S.~J.~Pennycock, \Nat {\bf
351}, 47 (1991).

\bibitem{channel_evidence} N.~F.~Heinig, R.~D.~Redwing, I Fei Tsu,
A.~Gurevich, J.~E.~Nordman, S.~E.~Babcock and D.~C.~Larbalestier,
\APL {\bf 69}(4), 577 (1996).

\bibitem{Buckel} Similar behavior in conventional, low-temperature
superconductors was described by W.~Buckel,
\emph{Superconductivity: Fundamentals and Applications} (VCH,
Wein\-heim, 1991) p.173.

\bibitem{Surzhenko01} A.~B.~Surzhenko, S.~Schauroth, M.~Zeisberger,
T.~Habisreuther, D.~Litzkendorf and W.~Gawalek, \SUST {\bf 14},
770 (2001).

\bibitem{creep} see, for example, A.~Gurevich and E.~H.~Brandt, \PRL {\bf
73}, 178 (1994).

\end{thebibliography}
\end{document}